# PENERAPAN TEKNOLOGI PENGOLAH CITRA DIGITAL DAN KOMPUTASI PADA PENGUKURAN DAN PENGUJIAN BERBAGAI PARAMETER BENANG

Andrian Wijayono[1] & Valentinus Galih Vidia Putra[1]

Textile Engineering Departement, Politeknik STTT Bandung, Indonesia[1]

**Abstrak:** Seiring perkembangan ilmu dan teknologi, penerapan berbagai proses komputasi pada bidang *science engineering* merupakan hal yang sedang dikembangkan, bahkan hingga saat ini. Saat ini, banyak penelitian yang menerapkan teknologi analisis citra pada bidang evaluasi dan identifikasi tekstil, salah satunya adalah benang. Teknik analisis citra komputer dapat digunakan untuk mengidentifikasi produk tekstil, khususnya produk tekstil berupa benang. Pada bab ini telah ditinjau berbagai metode dan penerapan teknologi *image processing* pada bidang identifikasi dan evaluasi berbagai parameter benang, bersamaan dengan tinjauan singkat perkembangan historis, serta kebutuhan akan penerapan metoda tersebut.

**Kata Kunci:** *benang, pengolah citra digital, evaluasi dan identifikasi, tekstil.*

# 1. PENDAHULUAN

Industri tekstil Dalam dunia tekstil, pemintalan didefinisikan sebagai suatu proses yang secara umum melibatkan tahapan peregangan (*drafting*), penggintiran (*twisting*) dan penggulungan (*winding*) untuk menghasilkan benang. Benang merupakan salah satu material yang menjadi bahan baku dalam berbagai penggunaan, baik sebagai material tekstil, maupun material non tekstil. Dalam dunia industri tekstil, kualitas dari berbagai parameter benang seperti nomor benang, diameter benang, antihan benang, dll merupakan hal yang penting untuk dijaga agar tetap sesuai dengan spesifikasi yang diharapkan.

Penerapan teknik pengolah citra digital pada rangkaian evaluasi dan identifikasi telah banyak diteliti oleh berbagai peneliti. Penerapan teknik tersebut pada berbagai bidang telah memungkinkan kita untuk dapat menilai, menyimpulkan dan mendapatkan suatu informasi pada suatu citra digital. Dengan menggunakan teknik pengolah citra, pengukuran berbagai parameter benang yang membutuhkan penilaian secara visual dapat dikuantifikasikan. Pada bab ini, akan dijelaskan lebih lanjut mengenai penerapan berbagai teknik pengolahan citra digital pada pengukuran dan identifikasi parameter benang oleh para peneliti.

# 2. PENERAPAN TEKNIK PENGOLAH CITRA PADA PENGUKURAN DIAMETER BENANG

V. Carvalho, dkk (2009) mengatakan bahwa diameter benang merupakan panjang sumbu lintang pada tepi benang satu terhadap lainnya, umumnya dinyatakan dalam bentuk nilai diameter rata-rata. Gambar 1.1 menunjukan skema jejari ($r$) dan sudut twist ($\vartheta$) dari suatu benang. Kretzschmar, Donmez dan Further (2009) menyatakan bahwa diameter benang bergantung pada jumlah serat yang berada pada penampang melintang benang, tingkat kehalusan serat, kerapatan serat, antihan pada benang dan struktur permukaan benang. Selain itu, Kretzschmar, Donmez dan Further (2009) juga mengatakan bahwa pengukuran diameter benang juga penting untuk mengestimasi parameter *cover factor, porosity* dari struktur kain tenun dan struktur kain rajut. Diameter juga dapat digunakan untuk menentukan parameter *stitch length* (panjang benang pada satu jeratan) dari struktur kain rajut.

Mengacu pada SNI 8213:2016 Tesktil – Benang Jahit, diameter benang dikategorikan sebagai salah satu syarat mutu pada benang jahit (benang jahit harus memiliki nilai maksimum diameter tertentu berdasarkan nomor benang dan antihan), selain itu, semakin kecil diameter benang, semakin sedikit jumlah serat pada daerah tampang lintang benang yang mengakibatkan kekuatan benang menurun dan hal ini menyebabkan putus benang semakin besar (Penava, 1997). Hal senada dikatakan oleh Prenzova (2000) yaitu kekuatan benang akan berkurang seiring dengan berkurangnya diameter benang yang dikarenakan berkurangnya serat pada daerah tampang lintang benang. Untuk menghitung besar diameter benang secara teori dapat digunakan rumusan Trommer (1995), $d_{yarn} = 0{,}035 \ldots 0{,}04\sqrt{Tex}$.

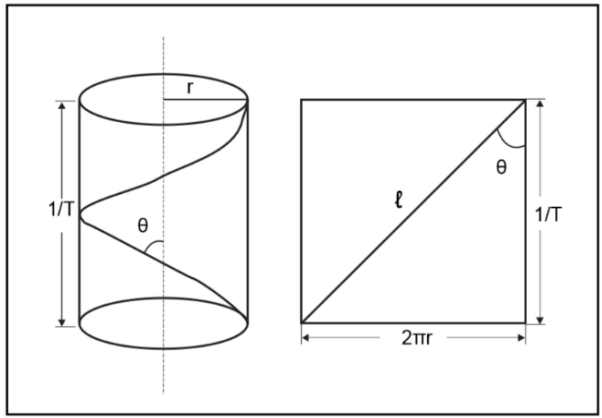

Sumber: nptel.ac.in
Gambar-1 Skema diameter dan antihan pada benang

Mahmoudi dan Oxenham (2002) telah menciptakan suatu alat pengukur ketebalan benang yang berbasis elektro-mekanis. Prinsip kerja alat pengukur tersebut dapat dilihat pada Gambar-2.

Menurut hasil penelitian Mahmoudi dan Oxenham (2002), kelemahan metoda pengukuran diameter benang dengan sistem mekanik adalah kemungkinan tidak dapat memberikan hasil pengukuran diameter yang valid dikarenakan adanya deformasi pada benang. Pada penelitian tersebut menunjukan bahwa pengaruh gaya luar pada metoda pengukuran diameter benang dengan sistem mekanik menyebabkan terjadi deformasi pada benang (semakin besar gaya luar yang diberikan pada benang, maka semakin kecil hasil pengukuran diameter benang pada metoda tersebut, dapat dilihat pada Tabel-1).

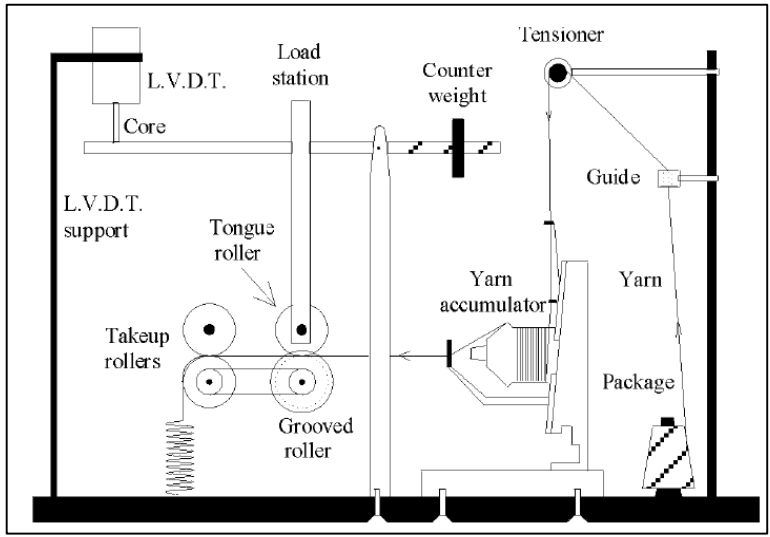

Gambar-2 Prinsip pengukuran ketebalan benang berbasis elektro-mekanis

Tabel-1 Ketebalan benang (diameter benang) pada pembebanan yang berbeda (W) untuk: benang worsted 37 Tex TPM 450.

| Pembebanan (gram) | Diameter (mm) | d0-d1 (mm) |
|---|---|---|
| Ketebalan semula* | 0.3191 | --------- |
| 0.2 | 0.2848 | 0.0343 |
| 0.5 | 0.2663 | 0.0528 |
| 1.0 | 0.2267 | 0.0924 |
| 5.0 | 0.2013 | 0.1178 |
| 10.0 | 0.1537 | 0.1654 |
| 20.0 | 0.1296 | 0.1895 |
| 30.0 | 0.1249 | 0.1942 |
| 40.0 | 0.1187 | 0.2004 |
| 50.0 | 0.1106 | 0.2085 |

*variabel ketebalan semula diukur dengan menggunakan metode pengukuran sistem optic

Sumber: Mahmoudi dan Oxenham (2002)

Lazlo, dkk (1994) mengatakan bahwa pengamatan terhadap diameter benang dengan menggunakan mikroskop merupakan metoda konvensional, namun teknologi *image processing* dapat memberikan hasil pengukuran diameter benang yang lebih akurat (tanpa adanya deformasi atau perubahan bentuk pada benang). Lazlo, dkk (1994) telah membuat sebuah alat pengukur diameter dan sudut antihan benang berbasis pengolahan citra digital dengan menggunakan perangkat mikroskop yang dimodifikasi menggunakan sebuah perangkat kamera CCD. Pada penelitian yang dilakukan Lazlo, dkk (1994), analisis citra diameter benang dilakukan dengan menentukan histogram level menggunakan aplikasi pengolah citra digital. Histogram level tersebut diambil dari nilai persebaran nilai *color map gray scale* yang ada pada *pixel* (satuan terkecil pada gambar atau citra digital). Skema instrumen pengukuran diameter benang yang telah dikembangkan oleh Lazlo (1994) dapat dilihat pada Gambar-3.

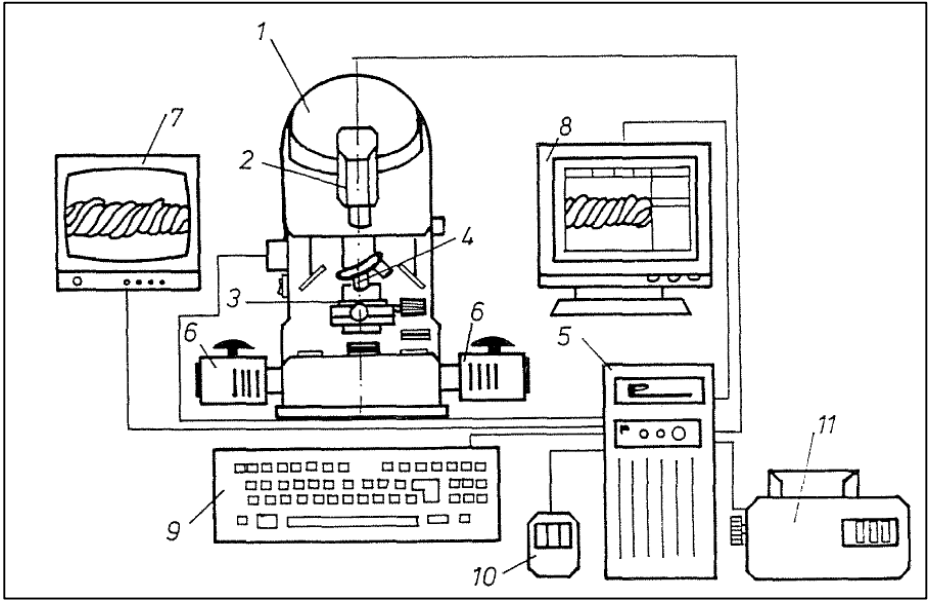

Gambar-3 Instrumen pengukur diameter dan sudut antihan yang dikembangkan oleh Lazlo (1994)

Pada alat yang dirancang oleh Lazlo (1994), penangkapan citra dilakukan dengan menggunakan sebuah perangkat mikroskop yang telah dipasang kamera CCD (pada titik 2) di bagian lensa okulernya. Kamera tersebut berfungsi untuk mengirimkan citra yang telah diproyeksikan oleh perangkat mikroskop pada komputer (5), yang pembesarannya dapat diatur sesuai dengan lensa objektifnya (pada titik 4). Contoh uji (pada titik 3) yang hendak diamati disimpan tegak lurus posisinya terhadap bidang pengamatan mikroskop.

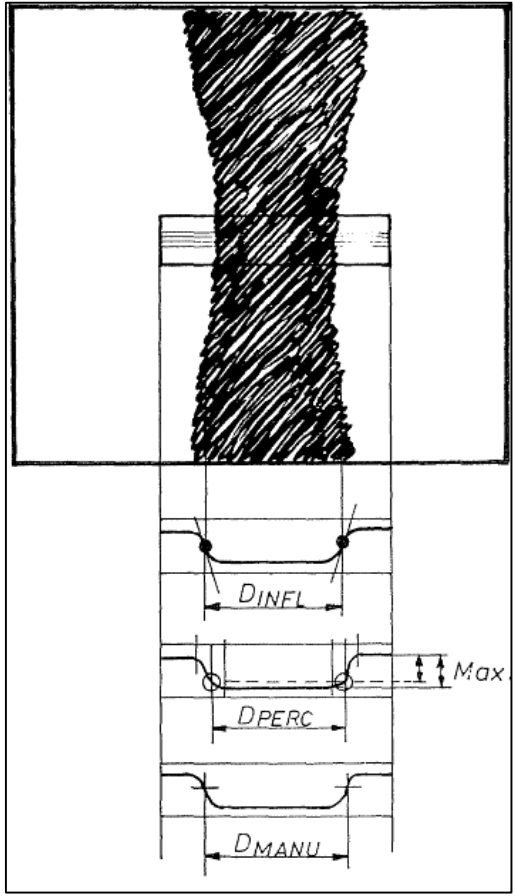

Gambar-4 Prinsip pembacaan nilai diameter pada alat pengukur diameter benang hasil rancangan Lazlo dkk (1994)

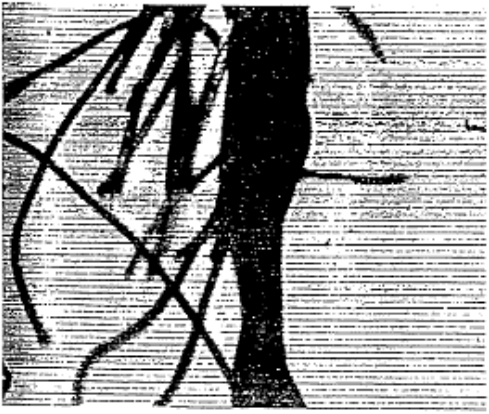
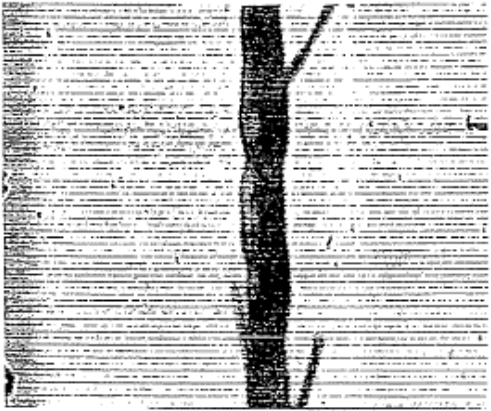
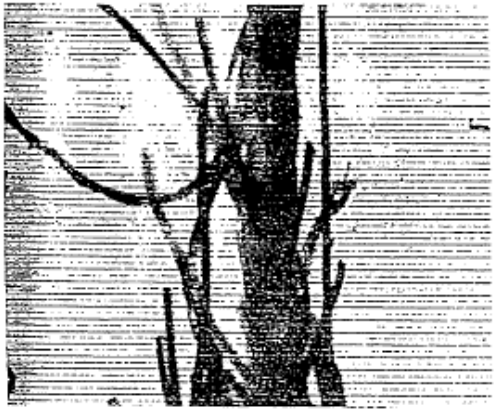

Gambar-5 Citra benang hasil tangkapan kamera CCD pada alat pengukur diameter benang (Lazlo, 1994)

Lazlo telah membuat pula sebuah perangkat lunak yang dapat digunakan untuk mengukur diameter benang secara otomatis, yang berprinsip kerja memanfaatkan histogram level pada suatu citra digital benang yang diamati. *Histogram level* tersebut akan diterjemahkan oleh perangkat lunak tersebut kedalam suatu bentuk kurva distribusi kumulatif nilai *pixel*. Pembacaan diameter benang diambil dengan dasar pembacaan variabel $D_{INFL}$ dan variabel $D_{PERC}$.

Prinsip pembacaan diameter benang alat tersebut dapat dilihat pada Gambar-4. Citra digital benang yang ada pada Gambar-5 merupakan hasil penangkapan citra digital yang dilakukan oleh kamera CCD, yang perbesarannya dapat diatur dengan mengatur lensa okuler pada perangkat mikroskop.

Wijayono (2017) juga telah membuat suatu perangkat pengukur diameter benang berbasis pengolah citra digital dengan menggunakan sebuah perangkat mikroskop digital sebagai penangkap citra digital. Gambar-6 menunjukan skema alat pengukur diameter benang yang diciptakan oleh Wijayono (2017).

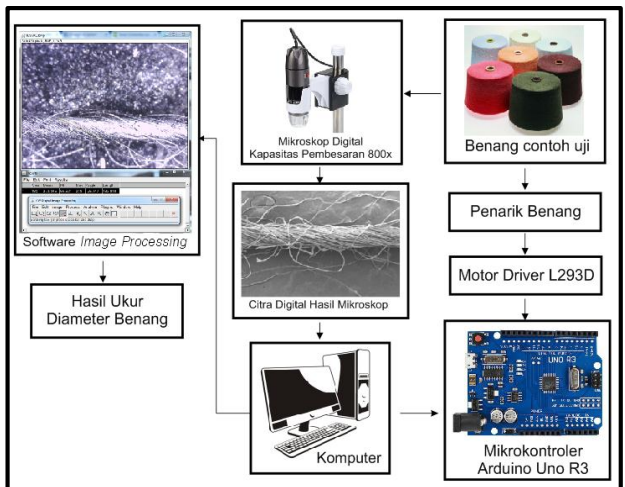

Gambar-6 Skema prinsip kerja alat pengukur diameter benang
(Wijayono, 2017)

Pada alat pengukur diameter benang tersebut, selain terdapat sistem penangkap citra (menggunakan perangkat mikroskop digital), terdapat pula mekanisme penguluran benang (berbasis mikrokontroler Arduino Uno). Perangkat lunak yang digunakan untuk menganalisis diameter benang dapat mengukur diameter benang dengan prinsip konversi satuan panjang *pixel* menjadi satuan metrik. Konversi satuan *pixel* menjadi satuan metrik dapat dilakukan jika benda yang tergambar pada citra digital diketahui panjang atau luasnya dalam satuan metrik. Gambar-7 merupakan contoh citra digital kawat tembaga berdiameter 0,5 mm yang dihasilkan oleh mikroskop digital.

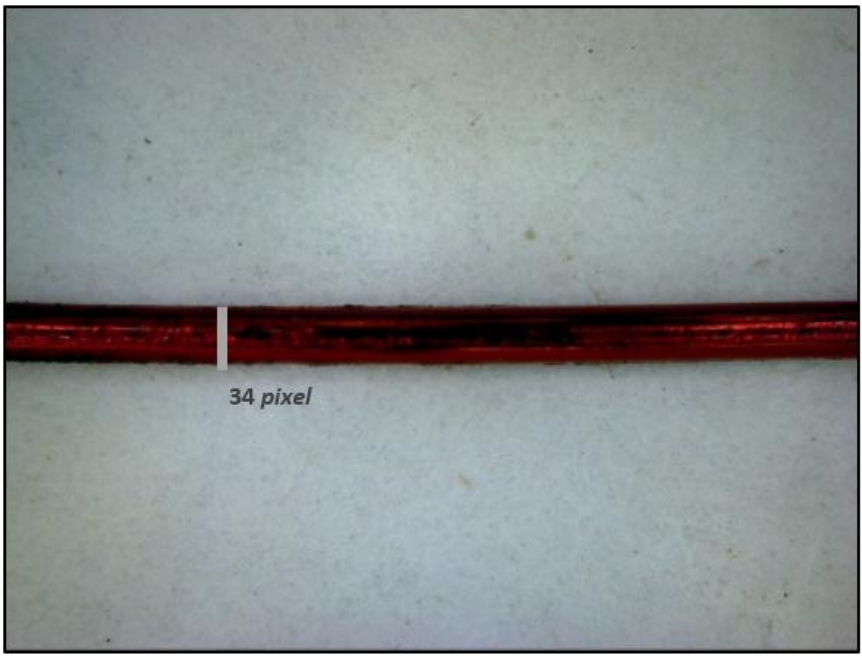

Gambar-7 Citra digital kawat tembaga berdiameter 0,5 mm

Dalam bidang *image processing*, satuan *pixel* merupakan satuan berdimensi [L]. Berdasarkan hasil penghitungan pada Gambar-7, jumlah *pixel* pada garis abu-abu ke arah penampang kawat adalah sebanyak 34 *pixel*. Metoda untuk mengkonversi tersebut dilakukan dengan merumuskan nilai $x$ dalam satuan

*pixel*/mm, nilai p dalam satuan *pixel* dan nilai $l$ dalam satuan mm sebagai berikut:

$$x = \frac{p}{l \, (mm)} \quad (1)$$

Dengan nilai p sebesar 34 *pixel* dan $l$ sebesar 0,5 mm, maka:

$$x = \frac{34 \, pixel}{0,5 \, mm} = 68 \, pixel/mm \quad (2)$$

Nilai konversi $x$ tersebut bersifat relatif terhadap pembesaran yang dilakukan oleh mikroskop digital. Apabila pembesaran (*zoom*) pada mikroskop digital berubah, maka nilai konversi tersebut juga berubah. Pada pembesaran yang sama (*zoom* tidak berubah), nilai konversi tersebut dapat digunakan untuk mencari nilai $l$ (mm) dari suatu benda yang tergambar pada suatu citra digital. Gambar-8 merupakan citra digital benang 40$^{'S}$ dengan nilai pembesaran yang sama dengan Gambar-7.

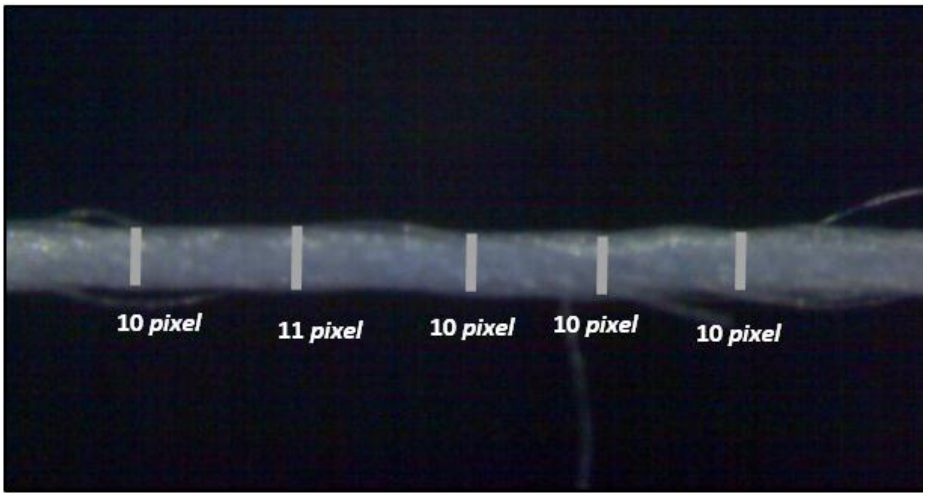

Gambar-8 Citra digital benang $Ne_1$ 40 poliester 100%

Apabila kita menghitung nilai $p$ (jumlah *pixel*) pada lima daerah benang pada Gambar-8, maka didapatkan nilai $p$ rata-rata sebesar 10,2 *pixel*. Dengan menggunakan angka konversi $x$ sebesar 68 *pixel*/mm, nilai diameter benang dapat diketahui dengan menghitung:

$$x = \frac{p}{l \ (mm)} \quad (3)$$

$$l \ (mm) = \frac{p}{x} = \frac{10{,}2 \ pixel}{68 \ pixel/mm} = 0{,}15 \ mm \quad (4)$$

kelemahan dari metode yang dirancang oleh Wijayono (2017) yaitu pengukuran benang belum memiliki fitur *image recognition*, yang memungkinkan pengukuran diameter benang secara otomatis untuk dapat dilakukan.

Adapun hasil analisis terhadap hasil pengukuran diameter dari alat pengukur yang dirancang oleh Wijayono (2017) dapat dilihat pada Gambar-9 berikut. Berdasarkan hasil analisa tersebut, alat pengukur diameter benang memiliki nilai persamaan yang mendekati teori, yang ditunjukan juga oleh grafik *trendline*-nya.

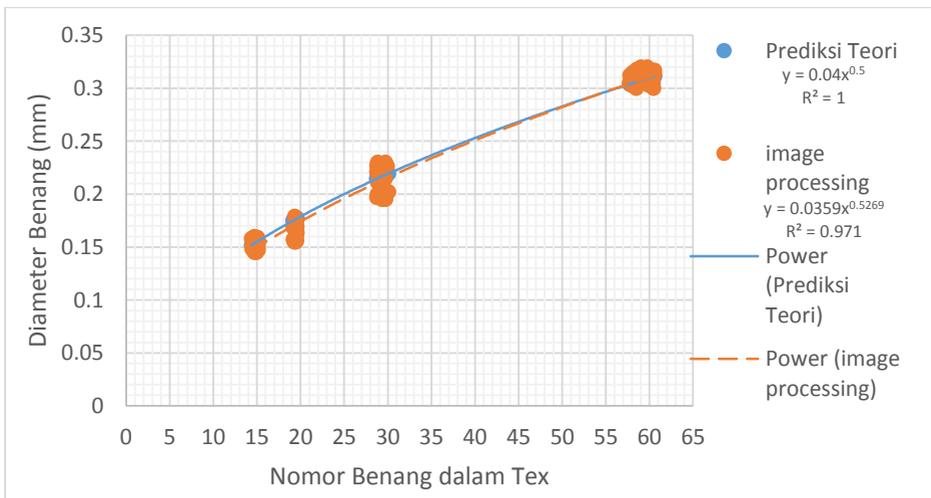

Gambar-9 Hasil analisis pengukuran diameter benang (Wijayono, 2017)

## 3. PENERAPAN TEKNIK PENGOLAH CITRA PADA PENGUKURAN ANTIHAN BENANG

Antihan adalah salah satu parameter penting pada benang. Antihan dapat menentukan berbagai karakteristik pada bahan seperti *hairiness, strength* dan nomor benang (Dewanto, Totong dan Putra, 2016). Menurut SNI ISO 17202:2010, jumlah antihan per satuan panjang tertentu atau *twist level* dapat dinyatakan dalam setiap meter (APM) atau setiap inci (API). Lawrence (2003) menyebutkan bahwa *twist level* dinyatakan dalam setiap inci dengan *turns per inch* (tpi) atau dalam satuan metrik yaitu *turns per meter* (tpm) dan *turns per centimeter* (tpcm).

Dalam rangka pengendalian mutu, jumlah antihan pada benang hasil produksi pemintalan dapat dilakukan sebagai salah satu langkah dalam rangka evaluasi benang. Hubungan antara antihan dan nomor benang dalam tex dijelaskan oleh Sema (2008) yaitu dengan semakin besar densitas panjang (tex) akan mengurangi besar antihan. Trommer (1995), Furter (2009) dan Lawrence (2010) menyatakan bahwa hubungan antara antihan dengan nomor benang dalam metrik (Nm) adalah berbanding lurus. Hubungan antihan dengan nomor benang menurut beberapa ahli dirumuskan sebagai berikut. Menurut Neckar (1971) antihan dirumuskan sebagai berikut

$$T = \alpha_m N_m^{0,6} \tag{5}$$

Dewanto, Totong dan Putra (2016) menyatakan rumus antihan pada mesin *ring spinning* sebagai berikut

$$Ts = \frac{n_{traveller}}{Vf} = \frac{n_{Bobbin}}{Vf} - \frac{1}{\pi D_B} \tag{5}$$

$$Tz = \frac{n_{traveller}}{Vf} = \frac{n_{Bobbin}}{Vf} + \frac{1}{\pi D_B} \tag{6}$$

$Ts$ dan $Tz$ adalah besarnya antihan pada benang dengan jenis *s-twist* dan *z-twist*, $n_{traveller}$ adalah kecepatan putaran traveler pada lintasan ring, $Vf$ adalah kecepatan pengantar benang (*front roll*) dan $D_B$ merupakan diameter bobbin.

Secara eksperimen, pengukuran jumlah antihan pada benang dilakukan dengan menggunakan metode *twist-untwist* pada alat *twist tester*. Seiring dengan berkembangnya teknologi, sekarang sudah ditemukan cara untuk mengukur sudut antihan dengan menggunakan teknologi *image processing*.

Pada penelitian Yildiz dkk (2015), telah diterapkan teknologi *image processing* pada pengukuran antihan benang. Pada penelitian tersebut, telah dianalisis antihan dari empat buah jenis benang jahit 100% poliester sebagai sampel uji. Gambar-10 menunjukan hasil tangkapan citra digital dari empat buah sampel benang sampel uji.

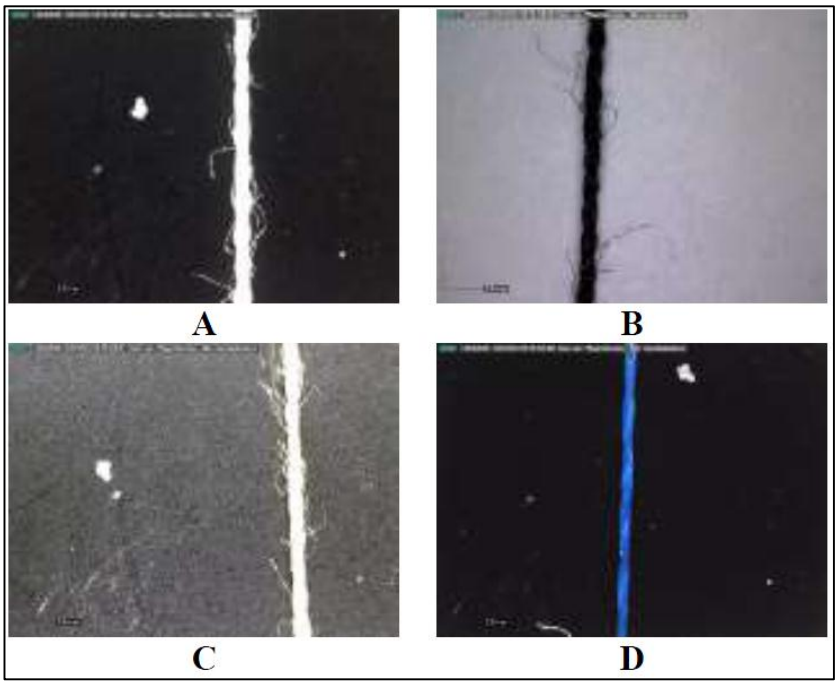

Gambar-10 Hasil tangkapan citra digital dari empat buah sampel benang sampel uji

Proses pengolahan citra pada penelitian Yildiz dkk (2015) dilakukan melalui tiga tahapan, yaitu tahap *preprocessing – I,* tahan *preprocessing – II* serta tahan analisis dan determinasi. Tahapan proses yang dilakukan pada penelitian ini dapat dilihat pada Gambar-11.

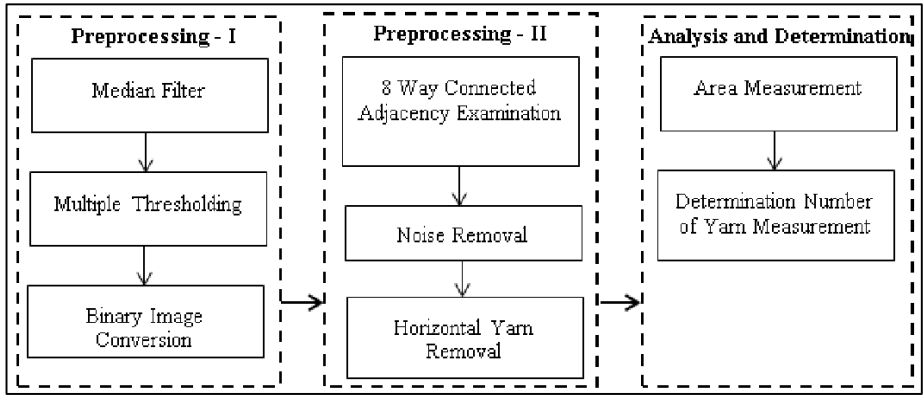

Gambar-11 Diagram blok tahapan

Pertama, setelah citra digital dalam skala abu (*grey scale*) didapatkan, citra digital selanjutnya dilakukan median filter. Median filter dapat membantu menghilangkan *noise* pada citra digital (sebelum dilakukan proses *noise removal*). Persamaan untuk median filter tersebut adalah sebagai berikut

$$x(m) = y_{med}(m) \qquad (7)$$

$$x(m) = Median[y(m-k), \ldots, y(m), \ldots, y(m+k)] \qquad (8)$$

$$x(m) = \begin{cases} y(m), if I y(m) - y_{med}(m) I < kQ(m) \\ else\ y_{med}(m) \end{cases} \qquad (9)$$

Proses *thresholding* digunakan dengan tujuan untuk membersihkan *noise* pada Gambar (Doğan, 2013). Pada proses *thresholding* tersebut, analisis citra digital dilakukan dengan menggunakan dua ambang batas nilai *pixel*, sehingga pada citra benang tersebut terbentuk suatu struktur yang dapat dianalisis. Nilai ambang bawah pada *thresholding* disebut sebagai "*mininum density threshold*" dan ambang atas pada *thresholding* disebut sebagai "*maximum density threshold*". Nilai *pixel* yang lebih kecil dari *mininum density threshold* dan lebih besar dari *maximum density threshold* diubah nilainya menjadi "0", daerah pixel tersebut akan muncul sebagai warna hitam pada citra. Gambar-12 (a) menunjukan sebuah matriks *gray level* sebuah citra dan Gambar-12 (b) menunjukan matriks citra digital setelah proses *thresholding*, dengan nilai *mininum density threshold* sebesar 80 dan nilai *maximum density threshold* sebesar 180.

| 246 | 255 | 250 | 245 | 253 | 255 | 255 | 255 |
|---|---|---|---|---|---|---|---|
| 167 | 192 | 207 | 221 | 245 | 255 | 255 | 255 |
| 128 | 127 | 136 | 157 | 194 | 224 | 246 | 255 |
| 175 | 154 | 134 | 123 | 127 | 139 | 165 | 186 |
| 165 | 163 | 162 | 175 | 156 | 134 | 123 | 120 |
| 64 | 98 | 130 | 163 | 163 | 173 | 179 | 163 |
| 10 | 19 | 41 | 65 | 96 | 132 | 159 | 188 |
| 4 | 9 | 13 | 16 | 23 | 35 | 57 | 93 |

(a) Gray Level Image Matrix

| 0 | 0 | 0 | 0 | 0 | 0 | 0 | 0 |
|---|---|---|---|---|---|---|---|
| 167 | 0 | 0 | 0 | 0 | 0 | 0 | 0 |
| 128 | 127 | 136 | 157 | 0 | 0 | 0 | 0 |
| 175 | 154 | 134 | 123 | 127 | 139 | 165 | 0 |
| 165 | 163 | 162 | 175 | 156 | 134 | 123 | 120 |
| 0 | 98 | 130 | 163 | 163 | 173 | 179 | 163 |
| 0 | 0 | 0 | 0 | 96 | 132 | 159 | 0 |
| 0 | 0 | 0 | 0 | 0 | 0 | 0 | 0 |

(b) Thresholded Image Matrix

Gambar-12 Ilustrasi proses *thresholding* pada suatu matrik citra *grey scale*

Setelah proses *thresholding*, kemudian citra digital diubah menjadi gambar *binary* pada proses *binary image conversion*. Gambar biner (*binary*) memiliki sebaran nilai *pixel* yang lebih sempit dibandingkan dengan *gray scale*. Hal tersebut dapat memudahkan proses pengecekan pengelompokan *pixel* pada gambar pada proses *"8 Way Connected Adjacancy Examination"*. Empat buah *pixel* atau lebih yang bernilai lebih dari satu akan dilabeli dengan *pixel* yang terkelompok atau tersambung (*adjacent*), sedangkan yang lainnya tidak. Gambar-13 menunjukan ilustrasi proses *8 Way Connected Adjacancy Examination* tersebut dilakukan.

Gambar-14 menunjukan mengenai cara pendeteksian diameter pada benang. Metoda tersebut menerapkan *histrogram level* untuk menentukan lebar benang pada citra digital. Posisi benang pada citra digital dapat diidentifikasi dengan melihat daerah puncak pada kurva *histogram level*.

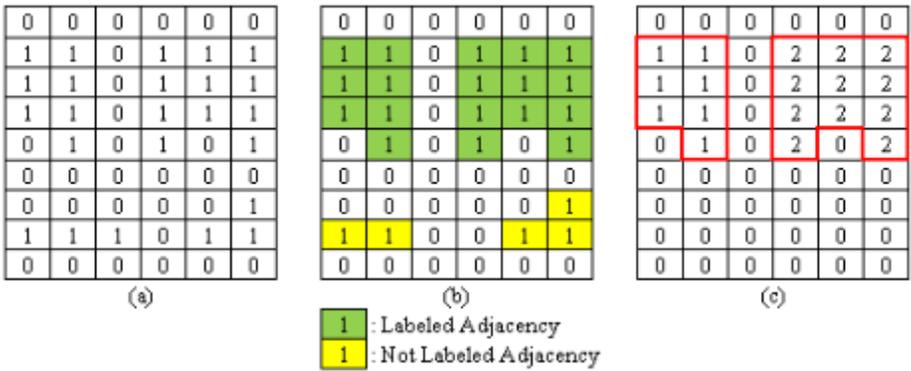

Gambar-13 Proses *8 Way Connected Adjacancy Examination*

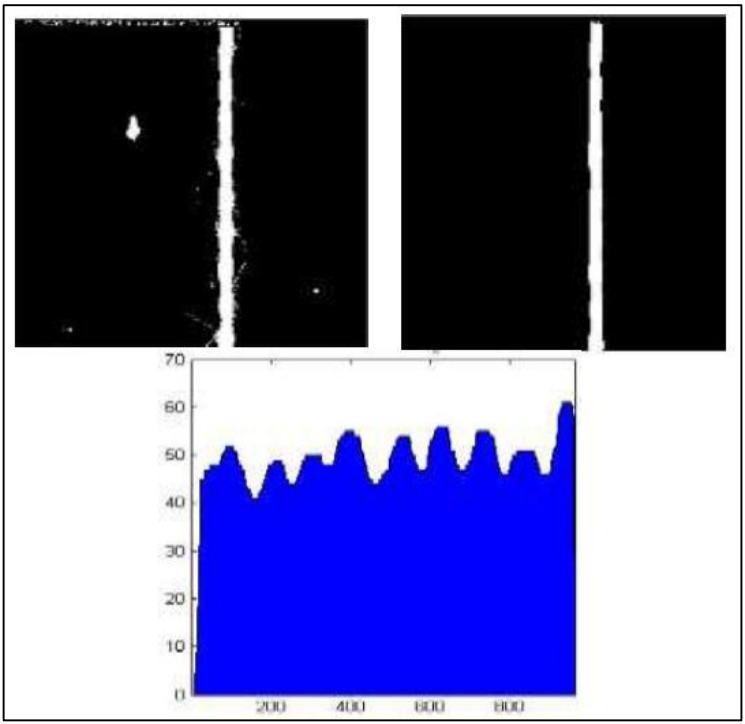

Gambar-14 Penerapan pengolah citra digital pada penentuan lebar (diameter) benang

Tabel-2 menunjukan hasil pengukuran dari alat yang dibuat oleh Yildiz dkk (2015). Berdasarkan hasil tersebut, dapat dilihat bahwa hasil pengukuran dari alat ukur antihan benang berbasis *image processing* dapat memberikan tingkat akurasi 85,7% hingga 94,6%.

Tabel-2 Hasil pengukuran antihan benang Yildiz dkk (2015)

| Sample Code | Twist value from twist tester (T/m) | Twist value from image processing (T/m) | Accuracy % |
|---|---|---|---|
| A | 374 | 391 | 94.6 |
| B | 437 | 434 | 99.1 |
| C | 426 | 478 | 89.1 |
| D | 186 | 217 | 85.7 |

Ozkaya (2010) juga telah menciptakan suatu instrumen alat pengukur antihan benang menggunakan teknik pengolahan citra digital. Perbedaan antara metode Ozkaya (2010) dan Yildiz (2015) adalah pada metoda pengolahan citra digitalnya. Pengolahan citra digital yang dilakukan oleh Ozkaya (2010) yaitu dengan menerapkan teknik spasial dan analisis gambar inti benang.

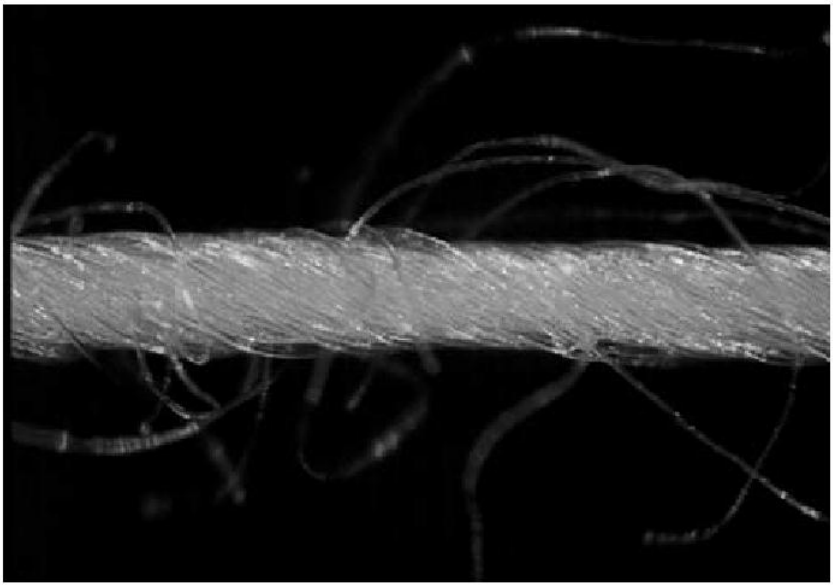

Gambar-15 Citra benang hasil tangkapan kamera CCD Hitachi HV C-10

Pada penelitiannya, Ozkaya (2010) menggunakan kamera CCD Hitachi HV C-10 untuk menangkap citra digital benang (contoh hasil tangkapan citra benang dapat dilihat pada Gambar-15). Citra benang yang telah ditangkap, selanjutnya akan dilakukan proses proyeksi horizontal untuk mendapatkan gambar benangnya saja (citra benang tanpa latar, selanjutnya gambar ini disebut sebagai *core yarn images* atau gambar inti benang). Proses tersebut dapat dilihat pada Gambar-16.

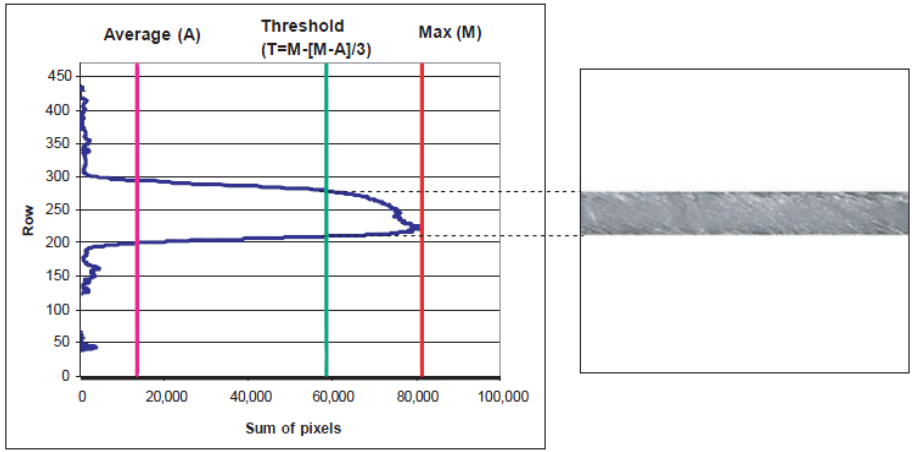

Gambar-16 Kiri: proses proyeksi horizontal untuk mendapatkan *core yarn image*, kanan: hasil proyeksi horizontal benang.

Untuk menganalisis antihan pada benang, Ozkaya (2010) menggunakan metoda *Fourier Transform* (FT) untuk mendapatkan proyeksi sudut antihan benang. Proses *Fourier Transform* (FT) untuk analisis benang tersebut dapat dilihat pada Gambar-17.

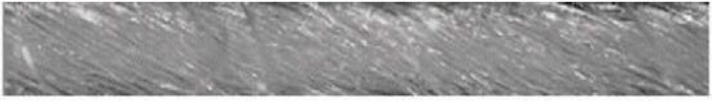
Original image

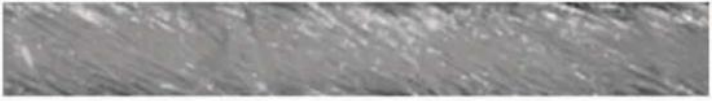
After low pass filtering

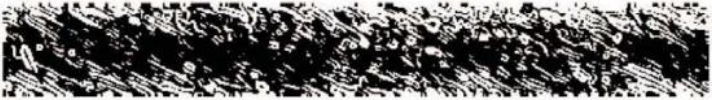
After applying a Sobel filter and thresholding

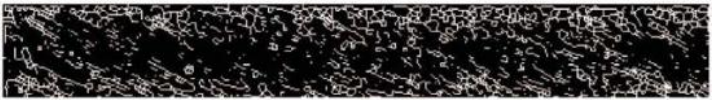
After skeletonising

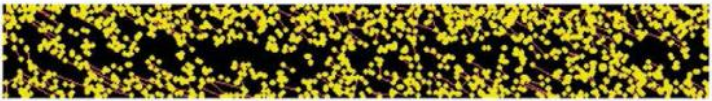
Lines extracted from skeleton contours

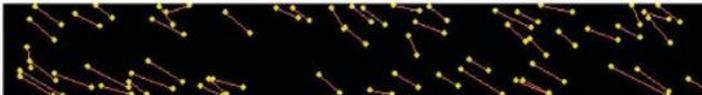
After merging close lines and filtering out the short ones

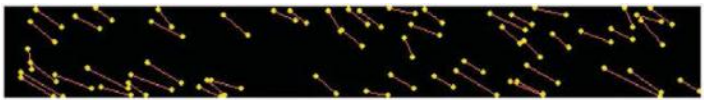
Lines in the dominant direction (S) are extracted

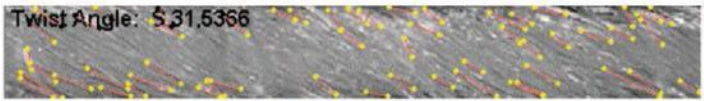
Output of the algorithm showing the weighed twist angle average

Gambar-17 Metode penentuan sudut antihan benang

# 4. PENERAPAN IMAGE PROCESSING PADA IDENTIFIKASI DISTRIBUSI KERAPATAN BULU (HAIR DENSITY DISTRIBUTION PROFILE) PADA BENANG

Konsep pengukuran bulu (*hairiness*) pada benang prinsipnya adalah mengukur jumlah atau total bulu yang ada pada benang. Perangkat pengukuran hairiness yang ada saat ini seperti USTER Hairiness Tester prinsipnya adalah menghitung jumlah ujung serat yang keluar dari permukaan benang dengan panjang sama dengan atau lebih besar dari 2mm (Ozkaya dkk, 2007). Berdasarkan hal tersebut, maka pengukuran tersebut tidak dapat merefleksikan perbedaan antara benang yang memiliki bulu yang panjang atau pendek, sehingga muncul suatu konsep penghitungan bulu yang dapat menghitung jumlah bulu panjang dan bulu pendek pada permukaan benang dengan menggunakan konsep *Hair Density Distribution Profile (HDDP)*.

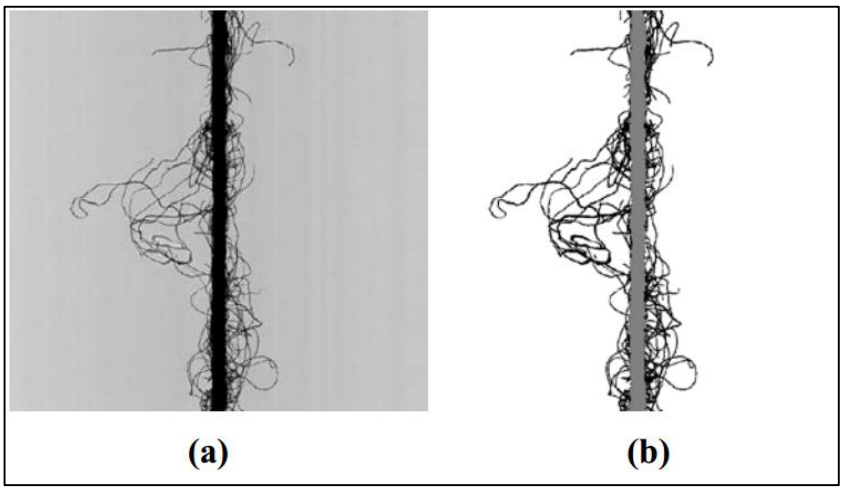

Gambar-18 Citra digital hasil tangkapan kamera CCD (a) dan citra hasil pengolahan algoritma (b)

Ozkaya dkk (2007) telah meneliti suatu metoda penentuan HDDP pada suatu benang. Pada penelitian yang dilakukan oleh Ozkaya dkk (2007), telah digunakan perangkat kamera CCD Dalsa Spark 2048-pixel dengan 100 mm

139

macro lens untuk menangkap citra digital benang yang diamati HDDP-nya pada pembesaran 1x. Citra digital benang pada Gambar-18 merupakan contoh citra yang telah ditangkap oleh kamera tersebut. Untuk memisahkan antara gambar penampang benang dengan gambar latar pada citra tangkapan CCD, telah digunakan suatu algoritma pengolahan citra digital yang telah diteliti sebelumnya oleh Ozkaya dkk (2005a). Pada Gambar-18 menunjukan suatu citra digital dari benang $Ne_1$ 10 65/35 poliester/katun pada penangkapan 70 scans/mm.

Pada penelitian Ozkaya dkk (2005b) telah membandingkan metoda HDDP berbasis *image processing* dengan HDDP rancangan USTER. Gambar-19 menunjukan hasil pebandingan antara pembacaan HDDP USTER dengan pembacaan yang telah dirancang oleh Ozkaya dkk (2005b).

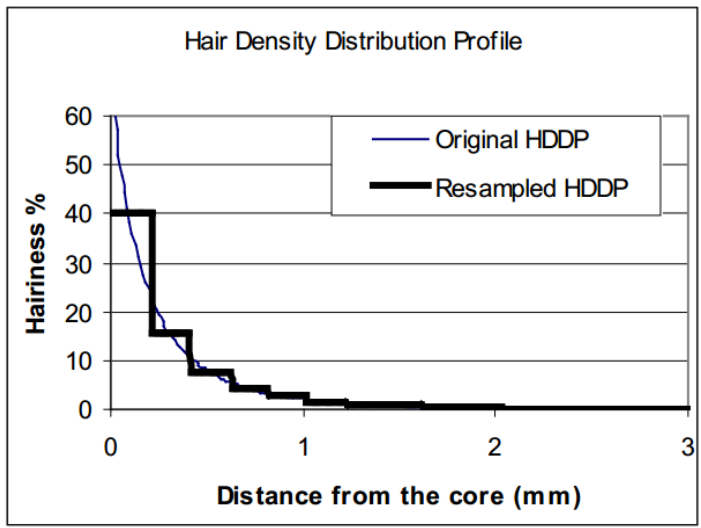

Gambar-19 Perbandingan HDDP USTER (*Original HDDP*) dengan HDDP (Ozkaya dkk, 2005)

Analisis lebih lanjut telah dilakukan oleh Ozkaya dkk (2007) untuk memodelkan dan mengidentifikasi HDDP pada benang. Gambar-20 menunjukan hasil identifikasi HDDP pada penelitian Ozkaya dkk (2007).

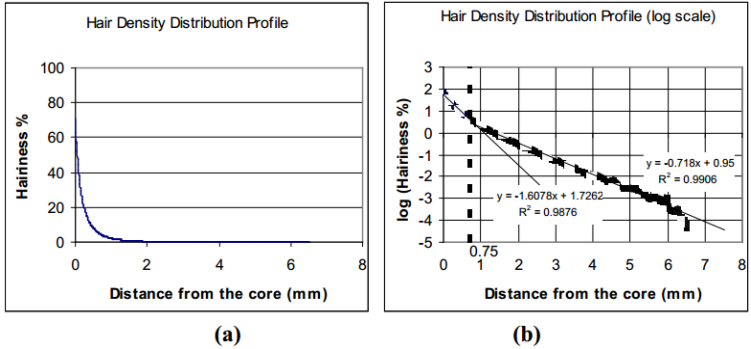

Gambar-20 Hasil analisis *hair density distribution profile* (a) dan hasil analisis *hair density distribution profile* pada kurva logaritma (b)

Berdasarkan analisis *fitting curve* pada penelitian Ozkaya dkk (2007), didapatkan nilai $R^2$ yang sangat baik (nilai $R^2$ lebih besar dari 0,97). Hasil analisis tersebut dapat dilihat pada Tabel-3.

Tabel-3 Hasil analisis *fitting curve* terhadap garis logaritmik $y = mx + b$

| Sample# | L≤0.75mm | | | L>0.75mm | | |
|---|---|---|---|---|---|---|
| | m | b | $R^2$ | m | b | $R^2$ |
| 1 | -1.67501 | 1.409163 | 0.984204 | -0.68557 | 0.526406 | 0.993075 |
| 2 | -1.58001 | 1.650767 | 0.990255 | -0.63263 | 0.776662 | 0.991058 |
| 3 | -1.51647 | 1.681013 | 0.990979 | -0.67315 | 0.909738 | 0.993908 |
| 4 | -1.66757 | 1.460766 | 0.987424 | -0.68911 | 0.542514 | 0.985387 |
| 5 | -1.64486 | 1.518171 | 0.990141 | -0.75078 | 0.675394 | 0.990886 |
| 6 | -1.58019 | 1.465919 | 0.992387 | -0.71122 | 0.616343 | 0.985147 |
| 7 | -1.44227 | 1.638258 | 0.988719 | -0.59056 | 0.842201 | 0.99238 |
| 8 | -1.50156 | 1.571486 | 0.990156 | -0.6195 | 0.728215 | 0.987933 |
| 9 | -1.558 | 1.396813 | 0.985396 | -0.60129 | 0.480479 | 0.980133 |
| 10 | -1.32814 | 1.220939 | 0.979529 | -0.42969 | 0.411879 | 0.988402 |
| 11 | -1.34831 | 1.397208 | 0.976136 | -0.42968 | 0.584741 | 0.989934 |
| 12 | -1.5059 | 1.678069 | 0.979368 | -0.43437 | 0.734776 | 0.98675 |
| 13 | -1.56168 | 1.21188 | 0.975801 | -0.54418 | 0.294518 | 0.986958 |
| 14 | -1.42761 | 1.659269 | 0.989584 | -0.54796 | 0.83444 | 0.989744 |
| 15 | -1.38377 | 1.55237 | 0.985808 | -0.51648 | 0.762061 | 0.992284 |
| 16 | -1.61861 | 1.483354 | 0.985894 | -0.596 | 0.547817 | 0.990131 |
| 17 | -1.46676 | 1.50621 | 0.984425 | -0.58938 | 0.725749 | 0.994914 |
| 18 | -1.46485 | 1.516771 | 0.985495 | -0.60036 | 0.758218 | 0.996364 |
| 19 | -1.6878 | 1.495208 | 0.986289 | -0.64342 | 0.500104 | 0.985446 |
| 20 | -1.37134 | 1.644449 | 0.990237 | -0.572 | 0.884903 | 0.99146 |
| 21 | -1.49901 | 1.464517 | 0.98711 | -0.61066 | 0.626112 | 0.987756 |
| 22 | -1.43648 | 1.362259 | 0.978574 | -0.56858 | 0.602718 | 0.990985 |

## 5. TEKNIK IDENTIFIKASI PARAMETER SLUB YARN BERBASIS ANALISIS CITRA

Benang *slub* dapat memberikan suatu kenampakan khusus pada permukaan kain, yang dapat digunakan pada berbagai aplikasi produk tekstil. Kenampakan khusus pada benang *slub* ditentukan oleh parameter benang tersebut, diantaranya panjang *slub*, jarak *slub*, amplitudo *slub*, dan periode pengulangan *slub* pada benang. Oleh karena itu, langkah pertama dalam pembuatan suatu kain *slub* (kain yang dibuat dari benang *slub*) adalah menganalisis parameter dari benang *slub*-nya (Pan dkk, 2011).

Metoda konvensional yang biasanya dilakukan untuk menganalisis parameter dari benang *slub* adalah dengan cara menghitung jumlah *slub* pada benang yang tergulung pada papan hitam, sehingga keandalan metoda tersebut bergantung pada keterampilan dan pengalaman penguji. Secara umum, pengulangan *slub* pada benang tidak akan terjadi pada panjang benang yang pendek, sehingga pengukuran parameter benang *slub* harus dilakukan minimal hingga pengulangan slub terjadi. Metode tersebut dinilai tidak efesien dan tidak efektif karena membutuhkan waktu yang lebih lama dalam hal analisisnya (Pan dkk, 2011).

Furter R () dalam penelitiannya telah menganalisis karakteristik benang *slub* dengan metode pengukuran menggunakan alat USTER. Alat yang digunakan oleh Furter (2005) berbasis sensor kapasitansi yang dapat mengukur parameter benang *slub*. Bian dkk (2006) juga telah mendeskripsikan suatu metode baru untuk mendeteksi parameter pada benang *slub* dengan menggunakan suatu alat akuisisi data.

Pada dasarnya, kedua penelitian yang telah dilakukan oleh Furter (2005) dan Bian (2006) merupakan penelitian yang berbasis sensor kapasitansi (Pan dkk, 2011). Menurut Pan dkk (2011), pengukuran parameter benang *slub* dengan menggunakan metoda tersebut belum dapat memberikan hasil yang benar-benar valid, mengingat nilai kapasitansi juga dapat dipengaruhi oleh jumlah serat yang berada pada penampang lintang benang, maka hasil kedua penelitian tersebut belum dapat menjabarkan parameter benang *slub* yang seharusnya ditinjau dari segi geometrisnya. Pada penelitian yang dilakukan oleh Pan dkk (2011), telah dilakukan suatu analisis citra untuk mengukur parameter pada benang *slub*.

Pan dkk (2011) memodelkan struktur dari benang *slub* seperti pada Gambar-21. $L_{bi}$ merupakan panjang benang normal; $N_{bi}$ merupakan nomor benang normal; $L_{si}$ merupakan panjang benang *slub*; $N_{si}$ merupakan nomor benang *slub* ($i$ = 1, 2, 3, … ).

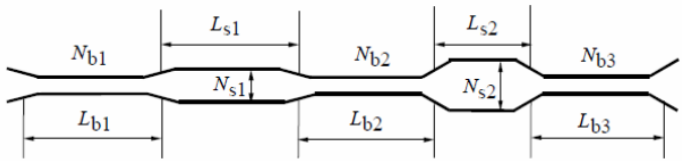

Gambar-21 Struktur benang *slub* (Pan dkk, 2011)

Pada penelitian yang dilakukan oleh Pan dkk (2011), terdapat beberapa langkah atau tahapan yang dilakukan untuk mengukur parameter benang *slub* dengan menggunakan prinsip pengolahan citra, diantaranya:

1. Proses untuk mendapatkan citra digital. Pada proses ini benang *slub* digulung pada suatu papan hitam dengan menggunakan YG381 *Yarn Evenness Tester*. Kerapatan benang yang digulung pada papan hitam berkisar antara tiga atau empat helai per centimeter. Penangkapan citra digital dilakukan dengan menggunakan sebuah perangkat *flat scanner* beresolusi 1200 dpi.
2. Proses *thresholding* citra digital. Proses ini bertujuan untuk menghilangkan *noise* pada citra digital hasil tangkapan *flat scanner*. Gambar-22 menunjukan citra digital hasil tangkapan *flat scanner* beresolusi tinggi, sedangkan Gambar-23 menunjukan histogram level dari citra digital yang telah ditangkap. Setelah proses *thresholding*, citra benang pada Gambar-22 akan berubah seperti yang ditunjukan pada Gambar-24. Bulu-bulu benang yang tertangkap pada citra dihilangkan, karena bulu bukan merupakan parameter benang *slub* yang diukur pada penelitian ini.

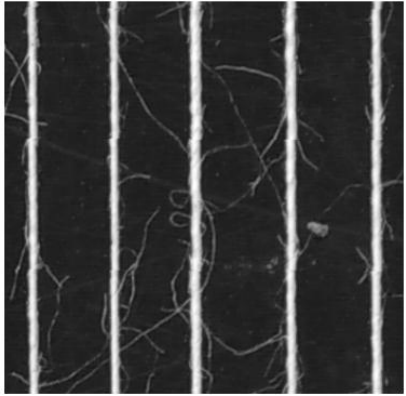

Gambar-22 Citra digital benang *slub* hasil tangkapan *flat scanner*

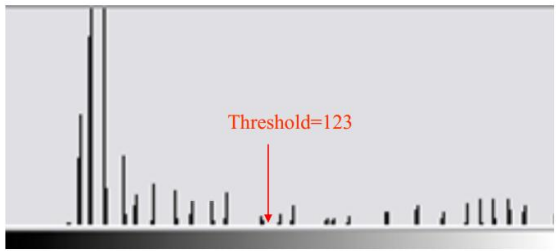

Gambar-23 Histogram level citra digital benang *slub*

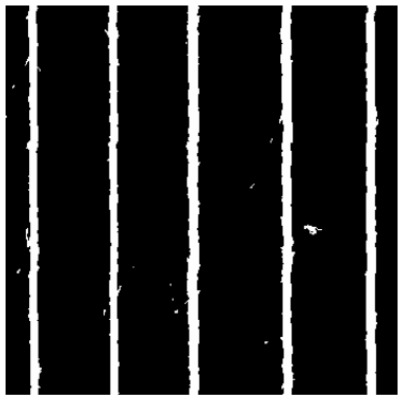

Gambar-24 Citra digital benang *slub* hasil proses *thresholding*

3. Penghilangkan objek kecil pada citra. Proses ini bertujuan untuk menghilangkan *pixel* berwarna putih pada citra digital hasil proses *thresholding*. Gambar-25 menunjukan hasil proses penghilangkan objek kecil pada citra yang telah dilakukan.
4. Memisahkan antara bagian *slub* dan bagian normal pada benang. Proses ini dilakukan dengan menggunakan fitur *edge detector* untuk mendeteksi pinggiran benang yang direpresentasikan pada citra digital. Hasil pengukuran *edge detector* dapat dilihat pada Gambar-26. Nilai yang tertera pada Gambar-26 menunjukan jarak antara kedua sisi pinggiran benang yang diukur berdasarkan jumlah *pixel* pada citra digital, maka lebar benang-benang pada citra digital dapat diketahui dengan mengukur setiap titik pada benang untuk mendapatkan nilai diameternya, kemudian jumlah titik yang memiliki masing-masing jumlah *pixel* tersebut dibuat kedalam bentuk histogram seperti pada Gambar-27.

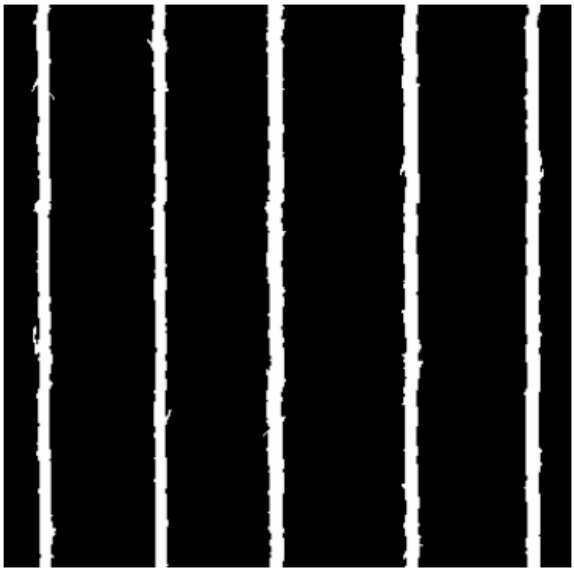

Gambar-25 Citra digital hasil yang telah bersih dari objek kecil

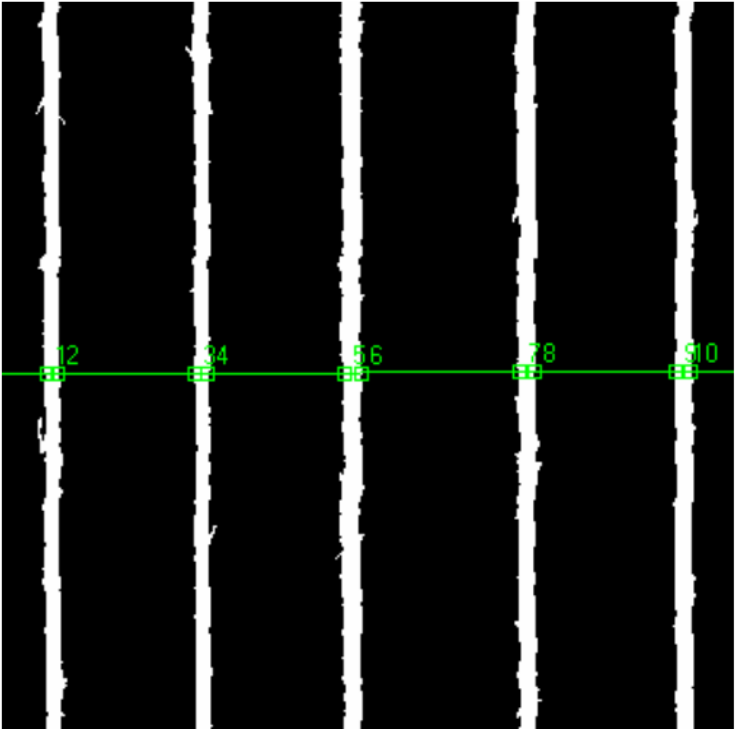

Gambar-26 Pembacaan *edge detector* pada citra digital

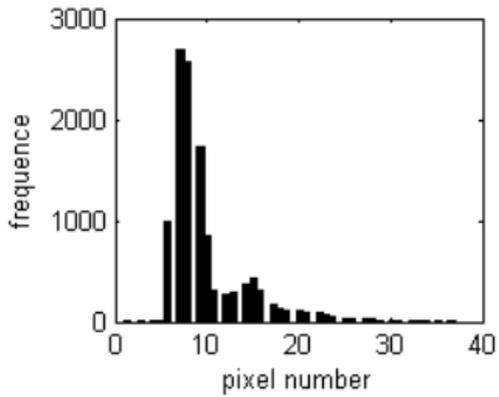

Gambar-27 Histogram *edge detector*

5. Membedakan antara ketidakrataan benang dengan *slub* benang. Pada proses pemintalan, terdapat kemungkinan akan adanya variasi diameter yang terjadi pada benang. Variasi tersebut mungkin saja dapat dideteksi sebagai *slub* pada proses *edge detector*. Oleh karena itu, panjang *slub* pada benang dibatasi minimal panjangnya adalah 20 milimeter, hal tersebut bertujuan untuk membedakan antara *slub* benang dengan ketidakrataan benang.

Pan dkk (2011) juga telah melakukan validasi eksperimen terhadap sampel benang *slub* berperiode dengan nomor benang $N_{bi}$ sebesar 14,5 tex. Hasil pembacaan panjang *slub,* amplitudo *slub*, dan jarak antar *slub* pada sampel benang *slub* dapat dilihat pada Tabel-3. Histogram panjang *slub* sampel uji dapat dilihat pada Gambar-28, serta Histogram amplitudo/pixel dapat dilihat pada Gambar-29.

Tabel-3 Hasil pengukuran parameter benang *slub* berperiode
(dengan $N_{bi}$ sebesar 14,5 tex)

| Panjang Slub (mm) | Amplitudo Slub (%) | Jarak Slub (mm) |
|---|---|---|
| 30 | 250 | 40 |
| 50 | 250 | 60 |

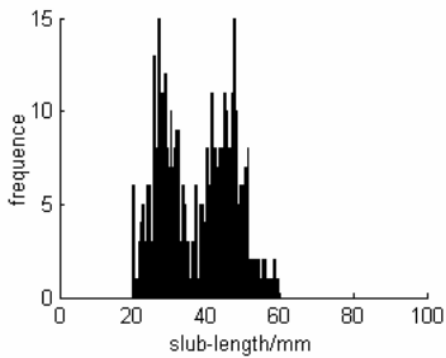

Gambar-28 Histogram panjang *slub*

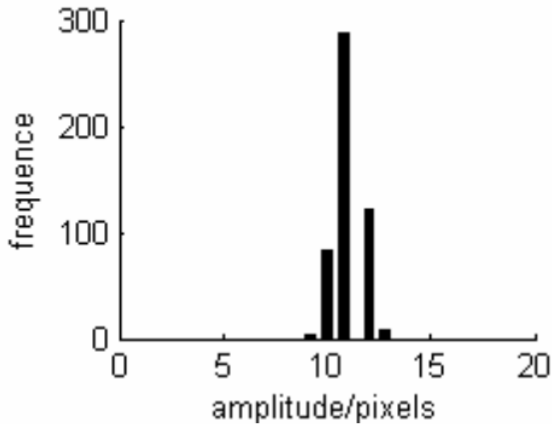

Gambar-29 Histogram amplitudo *slub*

## 6. PENGKLASIFIKASIAN DAN PENILAIAN CACAT SPLICING BENANG MENGGUNAKAN IMAGE PROCESSING

Menurut Gebald (1983), *Splicing* (proses penyambungan benang dengan menggunakan perangkat *spicer* atau penyambung benang) adalah suatu teknik untuk menyambungkan dua helai ujung benang. Cacat benang pada proses *winding* akan dihilangkan dengan suatu mekanisme *yarn clearing*. Benang yang memiliki cacat akan terdeteksi oleh perangkat *yarn clearer* dan akan dipotong untuk dibuang bagian cacatnya. Selanjutnya kedua ujung benang akan disambungkan kembali pada bagian proses *yarn splicing* yang umumnya menggunakan suatu perangkat *yarn splicer*.

Teknologi *splicing* di dunia industri telah berkembang sangat pesat seiring dengan perkembangan zaman. Terdapat berbagai macam teknik *splice* yang ada saat ini, yaitu *electrostatic splicing, mechanical splicing* dan *pneumatic splicing* (Isa dan Nagashi, 2005). Diantara ketiga metode *splicing* tersebut, *pneumatic splicing* merupakan metode yang paling popular digunakan di dunia industri.

Prinsipnya, proses *splicing* terdiri atas dua tahap, yaitu pembukaan antihan pada benang dan pemberian antihan kembali pada benang dengan menggunakan hembusan udara (Isa dan Nagashi, 2005). Kimura (1984) dan Mima (1984) menyatakan bahwa pada masing-masing proses pembukaan twist dan pemberian antihan, akan memiliki intensitas hembusan udara yang berbeda. Hembusan udara yang pertama akan menyebabkan ujung benang terbuka antihannya, kemudian hembusan udara kedua akan memiliki arah yang berlawanan dengan hembusan pembukaan antihan, yang akan menyebabkan pemberian antihan kembali pada benang.

Hingga saat ini, pengecekan mutu benang hasil penyambungan benang dan penilaian hasil pembukaan antihan masih dilakukan dengan menggunakan penilaian subjektif mata manusia (Issa dan Grutz, 1999). Kelemahan dari proses penilaian dengan menggunakan metode tersebut adalah membutuhkan waktu yang lama serta hasil yang didapatkan kemungkinan bervariasi karena faktor subjektifitas.

Pada penelitian yang dilakukan oleh Isa dan Nagahashi (2005), telah dilakukan suatu pendekatan untuk melakukan penilaian pada hasil pembukaan antihan (*opening*) dan hasil penyambungan benang (*splicing*) dengan menggunakan metoda pengolahan citra digital. Tabel-4 menunjukan pengklasifikasian mutu ujung benang *opening* pada proses penyambungan benang.

Jenis mutu ujung benang hasil proses *opening* tersebut kemudian diolah dengan menggunakan suatu algoritma pada perangkat lunak MATLAB. Isa dan Nagahashi (2005) mengimplementasikan metode *median filtering* 5 x 5 pada citra benang untuk mengekstraksi bentuk kontur ujung benang *opening*. Proses pengolahan citra tersebut dapat dilihat hasilnya pada Gambar-30. Citra benang yang telah dilakukan proses *median filtering,* selanjutnya akan dikonversi menjadi gambar hitam putih B/W. Objek-objek atau pixel kecil yang bukan merupakan bagian dari badan benang kemudian dihilangkan, sehingga akan dihasilkan gambar kontur dari benang dalam bentuk hitam putih.

Tabel-4 Klasifikasi mutu ujung benang hasil proses *opening* pada penyambungan benang

| Opening criterion | Grade | Opening shape |
|---|---|---|
| Length of the opening zone | optimum | |
| | average | |
| | short | |
| Untwist of the opening zone | optimum | |
| | partially | |
| | Partially overturn | |
| | Complete overturn | |
| | zero | |

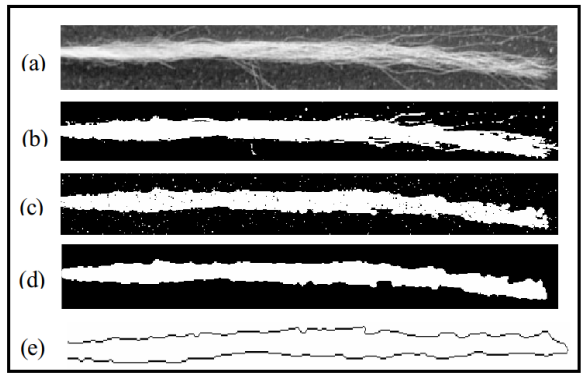

Gambar-30 Tahap awal pengolahan citra benang, (a) Citra digital asli benang, (b) Citra biner benang, (c) Citra biner benang yang masih memiliki *noise*, (d) Citra benang yang telah dilakukan proses *filter* dan (e) Citra kontur benang

Proses ekstraksi data yang dilakukan oleh Isa dan Nagahashi (2005) dapat dilihat pada Gambar-31. Variabel algoritma yang diterapkan pada proses tersebut antara lain:

- Y : Width of the parent yarn
- L : Length of the opening zone
- *Opening zone* dibedakan menjadi tiga buah daerah, yaitu sebagai berikut
    - Daerah kesatu (W1), adalah daerah *opening* yang dimulai dari titik *start point* hingga 5mm setelahnya.
    - Daerah kedua (W2), adalah daerah *opening* pada panjang antara 5mm hingga 10mm.
    - Daerah ketiga, adalah daerah yang lebih panjang dari 10mm.

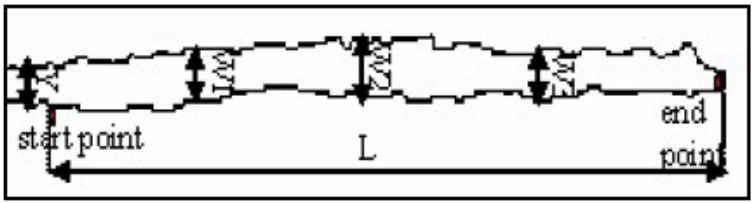

Gambar-31 Proses ekstraksi data pada citra benang hasil proses *opening*

Berdasarkan data yang diperoleh pada proses ekstraksi yang telah dilakukan sebelumnya, Isa dan Nagahashi (2005) mengklasifikasikan adanya cacat pada ujung benang *opening* sesuai dengan parameter berikut ini:

- Jika W1 lebih kecil atau sama dengan Y, maka diklasifikasikan sebagai *defect opening* (D).
- Jika W1 lebih besar dari Y dan panjang dari *opening zone* lebih kecil dari 5mm, maka *opening* diklasifikasikan sebagai *short length with good opening* (C1), atau dapat diklasifikasikan juga sebagai *partialy opening* (C2) didasarkan pada derajat *untwist* (tergantung pada perbandingan nilai W1 dan Y).

- Jika panjang *opening* L lebih besar dari 5 mm dan lebih kecil dari 10 mm, maka *opening* diklasifikasikan sebagai *medium length optimum opening* (B1), *partially opening* (B2), *partially overturn* (E), atau *complete overturn* (F).
- Jika panjang lebih besar dari 10mm, maka *opening* diklasifikasikan sebagai *optimum opening* (A) atau *partially overturn* (E) berdasarkan perbandingan antara W3 dan Y.

Gambar-32 menunjukan pohon pengklasifikasian *opening* sesuai dengan parameter yang telah ditetapkan oleh Isa dan Nagahashi (2005). Gambar-33 menunjukan grafik *acceptance level* pada proses pengklasifikasian yang telah dilakukan, untuk membedakan antara benang *opening* yang baik, buruk atau memenuhi syarat.

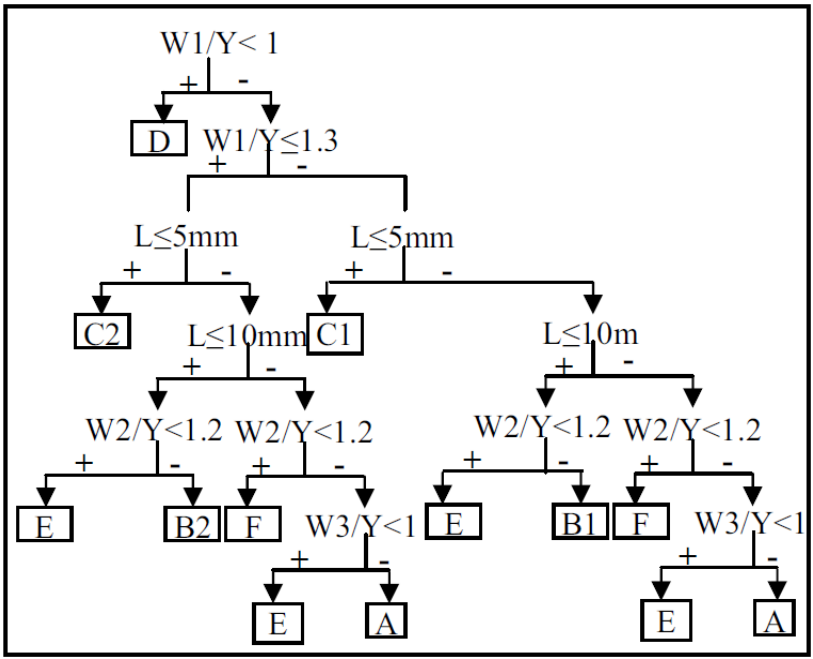

Gambar-32 Diagram pengklasifikasian benang *opening*

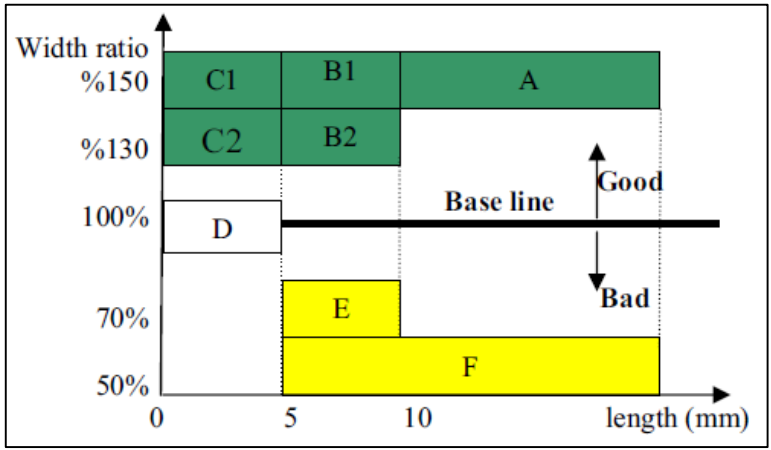

Gambar-33 Grafik penerimaan standar mutu benang *opening*

Tabel-5 Hasil pengklasifikasian mutu benang *opening* dengan menggunakan *image processing*

| Kriteria | N | Image Processing | | |
|---|---|---|---|---|
| | | Benar (C) | Salah (I) | Classification error (%) |
| A | 50 | 50 | 0 | 0 |
| B1 | 6 | 4 | 2 | 33 |
| B2 | 8 | 8 | 0 | 0 |
| C1 | 4 | 4 | 0 | 0 |
| C2 | 12 | 12 | 0 | 0 |
| D | 8 | 7 | 1 | 13 |
| E | 24 | 24 | 0 | 0 |
| F | 8 | 8 | 0 | 0 |
| Total | 120 | 117 | 3 | 6 |

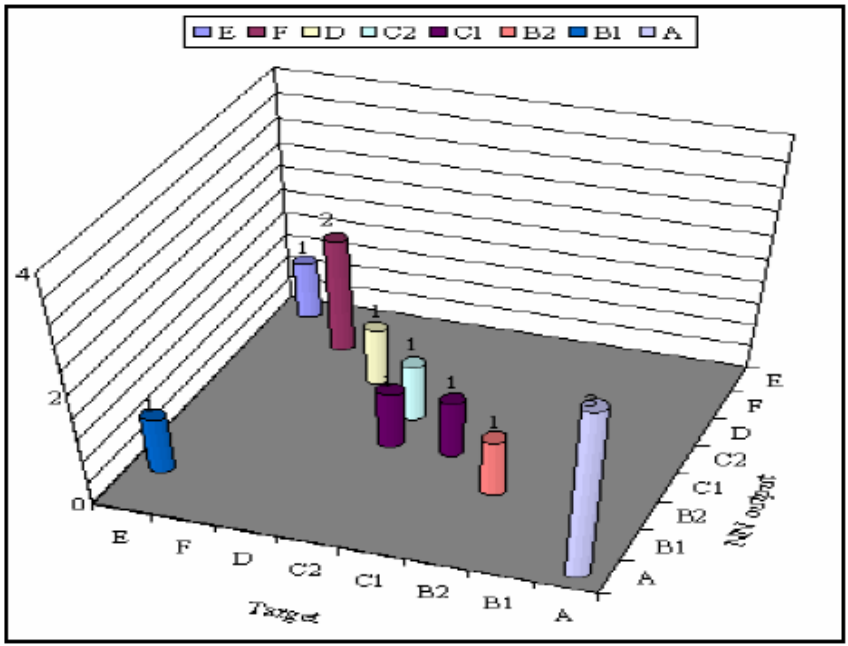

Gambar-34 Hasil pengklasifikasian mutu benang *opening*

Dengan jumlah sampel benang yang terbatas, Isa dan Nagahashi (2005) telah melakukan proses validasi eksperimen terhadap algoritma yang telah dirancang. Tabel-5 dan Gambar-34 menunjukan hasil pengklasifikasian mutu benang *opening* yang telah dilakukan oleh Isa dan Nagahashi (2005). *Classification error %* menyatakan persentase kesalahan deteksi pada proses pengklasifikasian mutu *opening*. Secara keseluruhan, persentase kesalahan pengklasifikasian hanya bernilai 6%.

## 7. ANALISIS PACKING DENSITY BENANG RING, COMPACT DAN VORTEX SPINNING DENGAN MENGGUNAKAN ANALISIS PENGOLAHAN CITRA

Menurut Kilic, Buyukbayrakter, Aydin dan Eski (2014), Kualitas suatu benang sangat dipengaruhi oleh karakteristik serat pada benang tersebut, seperti jenis serat, panjang serat, kehalusan serat, kekuatan serat, dll. Selain itu, susunan serat pada benang juga dapat mempengaruhi difat benang. *Packing density* menunjukan derajat susunan serat pada suatu benang, yang dihitung berdasarkan perbandingan antara luas daerah penampang serat terhadap luas daerah penampang lintang benang. Diameter benang, kekompakan benang, kontraksi benang, *porosity* benang dan volume benang secara langsung dipengaruhi oleh *packing density* benang. Selanjutnya, sistem teknologi pemintalan dapat mempengaruhi karakteristik benang, serta dapat mempengaruhi susunan serat pada struktur benang.

Kilic dkk (2014) telah melakukan penelitian untuk membandingkan *packing density* dari benang 100% Tencel LF yang diproduksi dari sistem pemintalan *ring, compact* dan *vortex spinning*. Nomor benang yang digunakan pada penelitian ini adalah 19,69 tex untuk setiap jenis benang dari berbagai sistem pemintalan. Konstanta antihan benang yang digunakan pada penelitian ini adalah $\propto_e = 3,7$ untuk benang dari sistem pemintalan *ring spinning* dan *compact spinning*, sedangkan *twist level* yang ditetapkan untuk benang *vortex* didasarkan pada parameter produksi yang sudah ditetapkan oleh Murata.

Penghitungan *packing density* yang dilakukan pada penelitian Kilic (2014), merupakan sistem penghitungan langsung (*direct method*). Pada metode ini, luasan serat pada penampang benang diukur berdasarkan luas penampang real serat pada citra digital. Analisis tersebut mensyaratkan pengolahan awal pada citra digital agar dapat memberikan hasil analisis *packing density* yang akurat, proses tersebut disebut sebagai proses *pretreated evaluation*.

Setelah proses *pretreated evaluation*, luasan serat diukur dan dibandingkan dengan luasan benang yang diperoleh. Pada penelitian Kilic (2014), bentuk penampang melintang dari benang diasumsikan dalam bentuk elips. Ukuran citra biner (*M* x *N*) dianggap sama dengan ukuran elips benang, sehingga luasan benang pada citra digital tersebut dapat ditulis sebagai $total\ yarn\ area = \frac{\pi MN}{4}$. Pada Gambar-35(c), garis warna merah menggambarkan daerah pinggiran benang. Serat-serat yang terletak diluar daerah benang (garis warna merah) akan diabaikan dan tidak akan dihitung dalam luasan total serat. Gambar-35 menunjukan langkah-langkah dari analisis citra dan hasil prediksi daerah benang.

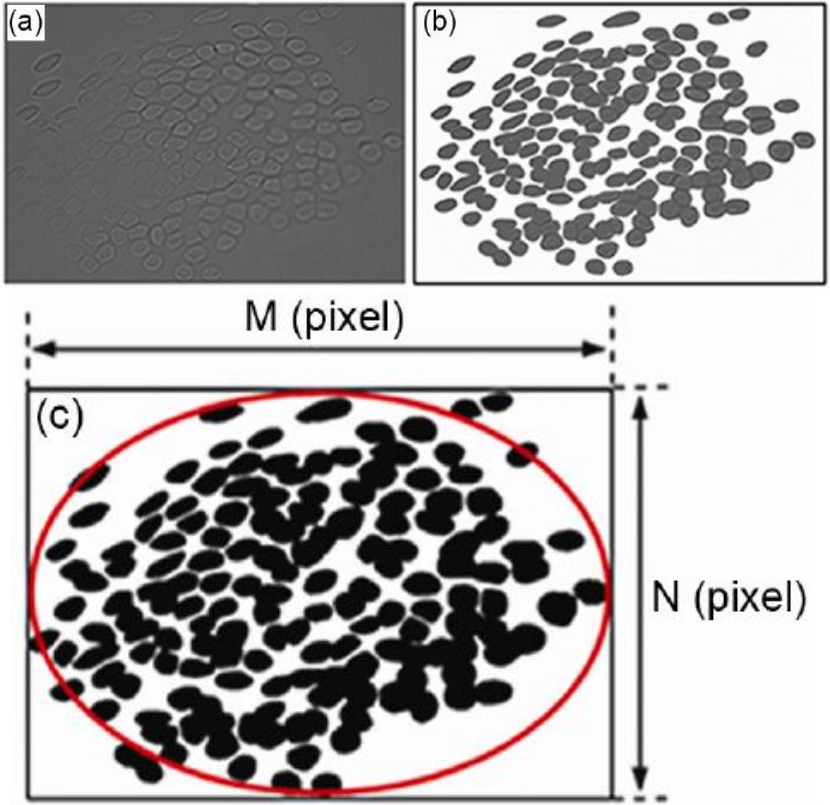

Gambar-35 Analisis citra digital pada pengukuran *packing density* benang

Citra penampang lintang benang ditangkap menggunakan perangkat SEM, kemudian citra digital dianalisis dan dihitung *packing density*-nya. Gambar-36 menunjukan citra digital SEM dari benang *ring, compact* dan *vortex spinning*.

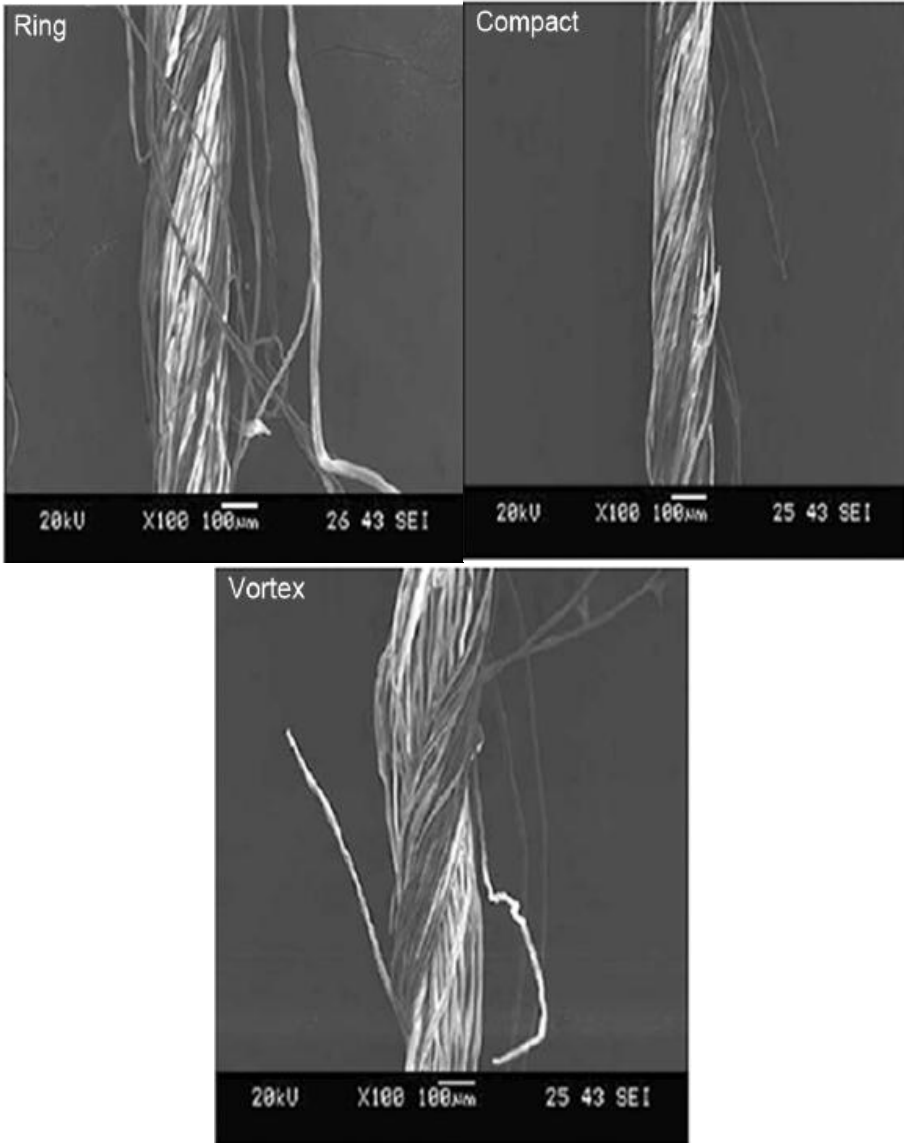

Gambar-36 Citra digital penampang membujur benang *ring, compact* dan *vortex spinning*

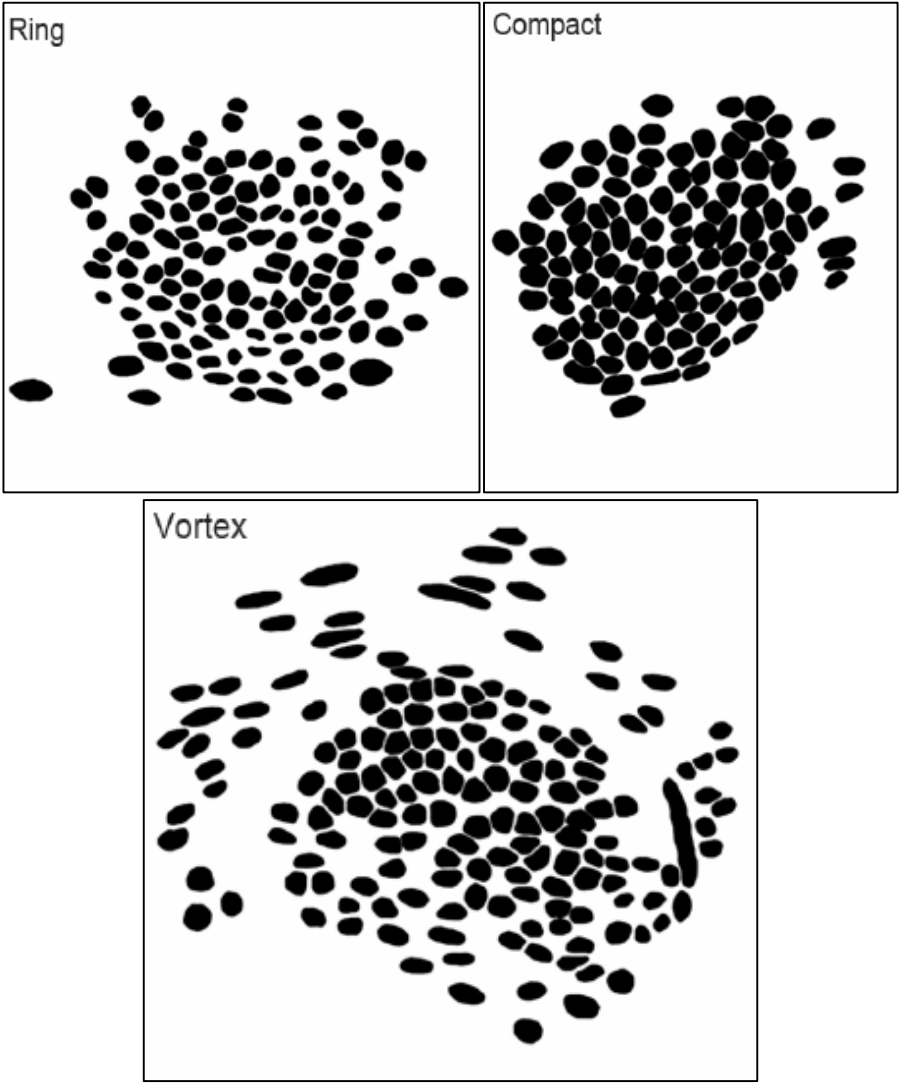

Gambar-37 Citra digital penampang melintang benang *ring, compact* dan *vortex spinning*

Tabel-6 menunjukan perbandingan nilai *packing density* dari masing-masing sistem pemintalan. Pengujian statistik pada data telah dilakukan untuk melihat pengaruh sistem pemintalan terhadap *packing density*. Hasil pengujian ANOVA dapat dilihat pada Tabel-7.

Tabel-6 Hasil pengukuran *packing density* benang pada penelitian Kilic (2014)

| Sistem pemintalan | *Density* (D) g/cm$^3$ | Diameter (2D$\oslash$) mm | *Packing Density* % |
|---|---|---|---|
| *Ring* | 0,66 | 0,194 | 38,01 |
| *Compact* | 0,67 | 0,194 | 41,91 |
| *Vortex* | 0,58 | 0,208 | 23,86 |

Tabel-7 Hasil analisis ANOVA *packing density* benang

| Source | *Sum of square* | Degree of freedom (df) | *Mean suqare* | F | Sig. |
|---|---|---|---|---|---|
| *Corrected model* | 902,195 | 2 | 451,098 | 20,406 | 0,000 |
| *Intercept* | 17951,864 | 1 | 17951,864 | 812,091 | 0,000 |
| *Spinning system* | 902,195 | 2 | 451,098 | 20,046 | 0,000 |
| *Error* | 265,269 | 12 | 22,106 | | |
| *Total* | 19119,328 | 15 | | | |
| *Corrected total* | 1167,464 | 14 | | | |
| *\*R squared = 0,773; Adjusted R square = 0,735* | | | | | |

Gambar-38 merupakan grafik interval kepercayaan ($\propto = 0,95$) untuk ketiga data benang dari masing-masing sistem pemintalan. Dapat dilihat bahwa nilai *packing density*, *density* dan diameter benang *vortex spinning* berada diluar daerah interval nilai dari benang *ring* dan *compact*.

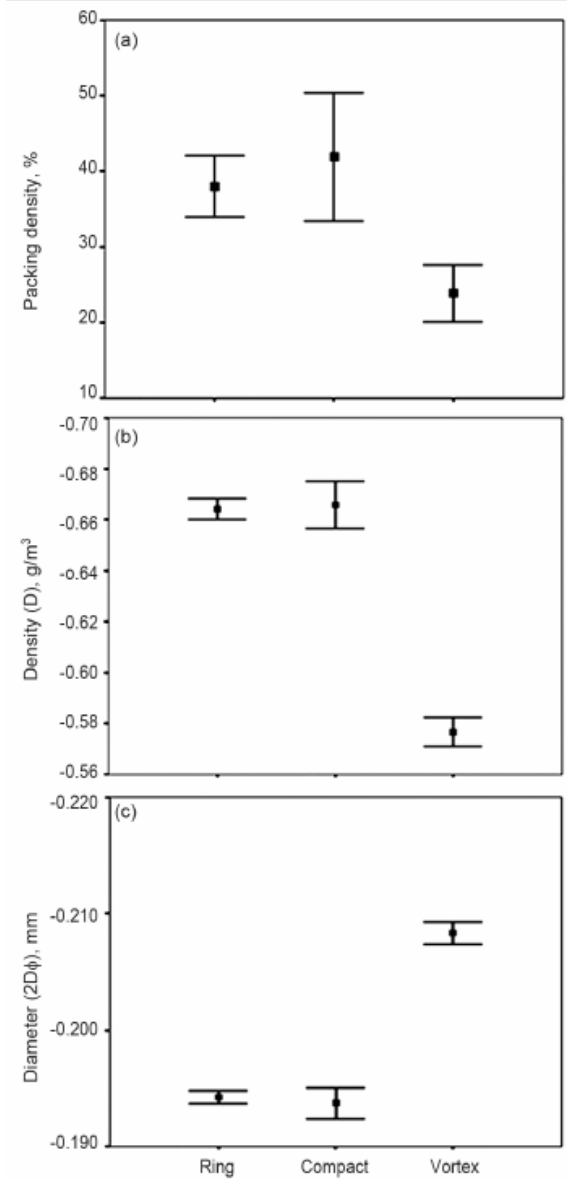

Gambar-38 *Packing density*, *density* dan diameter benang *ring, compact* dan *vortex spinning* pada interval kepercayaan 95%

Tabel-8 Analisis kesamaan nilai *packing density* antara benang sistem pemintalan *ring spinning, compact spinning* dan *vortex spinning*

| Spinning system (I) | (J) | Mean difference (I-J) | Standard error | Sig. |
|---|---|---|---|---|
| Ring | Compact | -3,896 | 2,974 | 0,215 |
| | Vortex | 14,154* | 2,974 | 0,000 |
| Compact | Ring | 3,896 | 2,974 | 0,215 |
| | Vortex | 18,050* | 2,974 | 0,000 |
| Vortex | Ring | -14,154* | 2,974 | 0,000 |
| | Compact | -18.050* | 2,974 | 0,000 |

*Mean difference dihitung pada level signifikansi 0,05

Berdasarkan data pada Tabel-8, dapat disimpulkan bahwa benang dari sistem pemintalan *ring spinning* dan *compact spinning* memiliki nilai yang tidak berbeda secara signifikan pada interval kepercayaan 95%. Pada interval kepercayaan 95%, *packing density* benang dari sistem pemintalan *vortex spinning* memiliki nilai yang berbeda dengan *packing density* benang sistem pemintalan lainnya.

## 8. PENERAPAN IMAGE PROCESSING UNTUK EVALUASI BENANG TEKSTUR FALSE-TWIST

Ghaderpanah, dkk (2014) telah mengembangkan suatu metode baru untuk mengukur parameter *crimp* pada benang tekstur dengan menggunakan *computer vision* dan *image processing*. Menurut Ghaderpanah, dkk (2014), penelitian tersebut telah berhasil memberikan hasil pengujian yang lebih akurat dan lebih cepat untuk menentukan parameter benang tekstur.

Benang tekstur merupakan suatu benang yang dihasilkan melalui proses *texturizing*. Menurut Ghaderpanah, dkk (2014), pemberian tekstur dimaksudkan untuk menciptakan deformasi permanen dari benang filamen, yang menciptakan suatu sifat yang berbeda pada benang. Secara umum texturizing dilakukan dengan salah satu dari tiga metode ini, yaitu secara mekanik, termal-mekanik atau kimia-mekanik. *Texturizing* dengan metode *false twist* adalah metode yang paling banyak digunakan dan praktis untuk menghasilkan benang tekstur. Terdapat berbagai faktor yang dapat mempengaruhi bentuk dan sifat crimp dari benang tekstur, diantaranya adalah struktur kimia filamen, nomor benang, jumlah filamen, bentuk luas permukaan dan faktor yang berkaitan dengan sistem seperti sistem gerak benang, jenis pemanas, panjang pemanas dan suhu pemanas.

Millman dkk (2001) telah memperkenalkan suatu sistem yang dapat mengukur parameter benang tekstur secara cepat dan akurat, serta tanpa terjadi kontak pada benang yang dapat mempengaruhi hasil pengukuran parameter benang. Sistem tersebut dapat mendeteksi derajat kemiringan filament pada benang, serta memiliki sensitifitas yang baik untuk mengukur perubahan diameter pada benang. Beberapa peneliti juga telah menggunakan beberapa metoda pengukuran *package density* benang, ketebalan kain serta jumlah kadar air pada benang tekstur untuk mengevaluasi *bulky* dan *stitch density*-nya.

Tujuan utama dari *texturizing* pada *air jet* adalah untuk menghasilkan benang yang memiliki *bulky*. *Bulk* dan *density of loops* merupakan faktor yang penting pada pengendalian mutu benang tekstur (Sengupta dkk, 1990).

Banyak proyek penelitian telah dilakukan untuk menyajikan metode baru untuk penentuan *crimp*, terutama konsentrasi *crimp* pada benang tekstur. Dengan munculnya komputer modern dan alat pemrograman baru, penggunaan *computer vision* untuk pengendalian mutu dan produk uji juga turut berkembang (Yousefzadeh dkk, 2005). Struktur benang *false twist* mirip dengan bentuk fraktal. Oleh karena itu, beberapa peneliti mengevaluasi *crimp* dari benang bertekstur berdasarkan geometri fraktal. Dalam penelitian Ghaderpanah, dkk (2014), metode baru untuk menghitung jumlah *crimp* filamen tunggal telah dikembangkan. Efek filamen dalam benang tekstur *false twist* dalam metode baru tersebut juga telah dipelajari.

Pada penelitian yang dilakukan oleh Ghaderpanah, dkk (2014), citra digital penampang membujur benang tekstur ditangkap dengan menggunakan perangkat *digital microscopic dino* (AU-351). Perangkat lunak MATLAB telah digunakan sebagai alat pengolah citra digital pada penelitian tersebut. Citra digital kemudian diubah menjadi bentuk hitam putih dengan menggunakan fitur im2bw pada MATLAB. Metode Otsu *thresholding* digunakan untuk memisahkan antara benang dengan latarnya. Proses tersebut akan menghasilkan suatu citra digital yang terdiri atas daerah putih dan hitam, daerah putih merepresentasikan filamen, sedangkan daerah hitam merepresentasikan latar.

Setelah didapatkan gambar dalam bentuk citra biner, kemudian dilakukan citra diolah dengan menggunakan operasi *thinner.* Citra biner dan citra hasil proses *thinner* dapat dilihat pada Gambar-39.

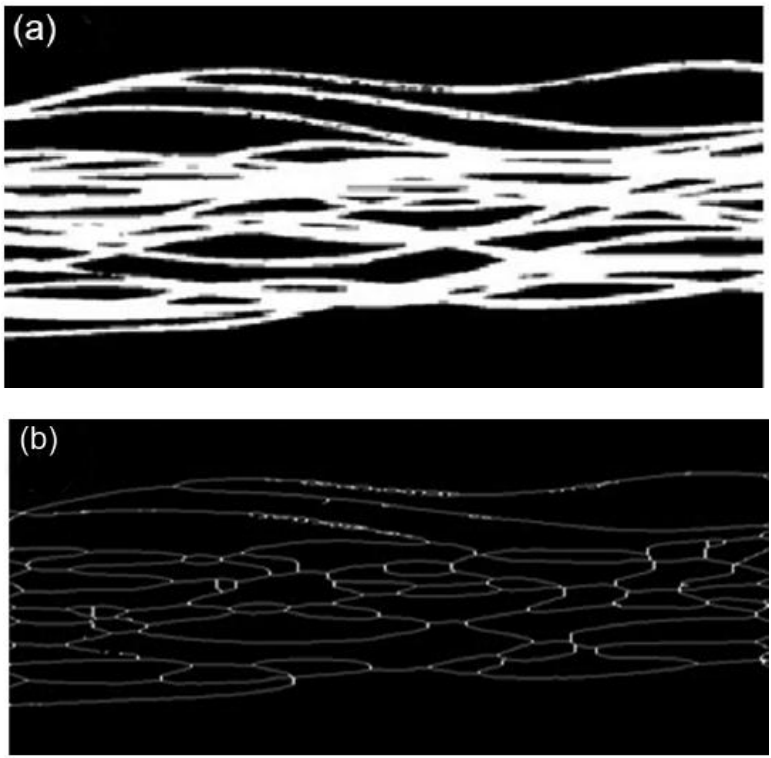

Gambar-39 Diagram skema dari benang tekstur *false twist* (a) dan citra hasil proses *thinning* (b)

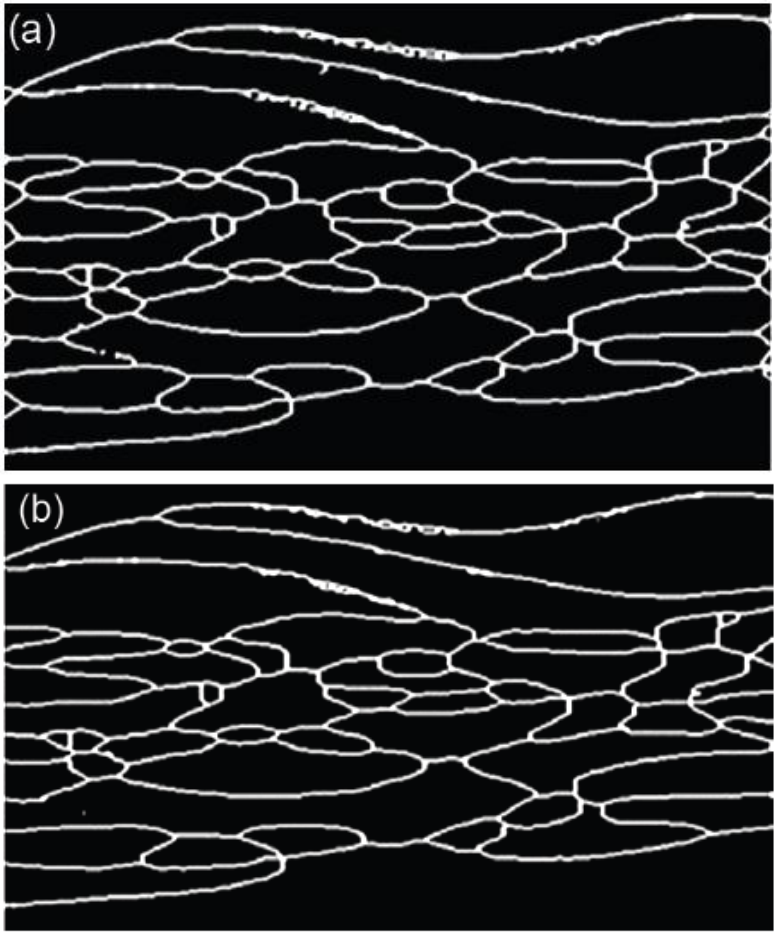

Gambar-40 Diagram skema dari benang tekstur *false twist* sebelum (a) dan setelah proses *corrective procedure* (b)

Serat pada benang dan ujung serat yang termasuk pada definisi *hairiness* akan ikut terproses pada pengolahan *thinning.* Ujung serat atau bulu tersebut harus dihilangkan karena bukan merupakan bagian dari bagian badan benang yang diukur parameternya. Proses *corrective procedure* dilakukan untuk membuang bagian ujung serat tersebut. Hasil proses *corrective procedure* dapat dilihat pada Gambar-40.

Untuk mengukur sudut kemiringan serat pada benang, dilakukan suatu metode *fiber tracing*. Metode tersebut akan mengukur sudut kemiringan benang dengan metode 8 tetangga atau 4 tetangga. Sistem koordinat 8 tetangga dan 4 tetangga dapat dilihat pada Gambar-41.

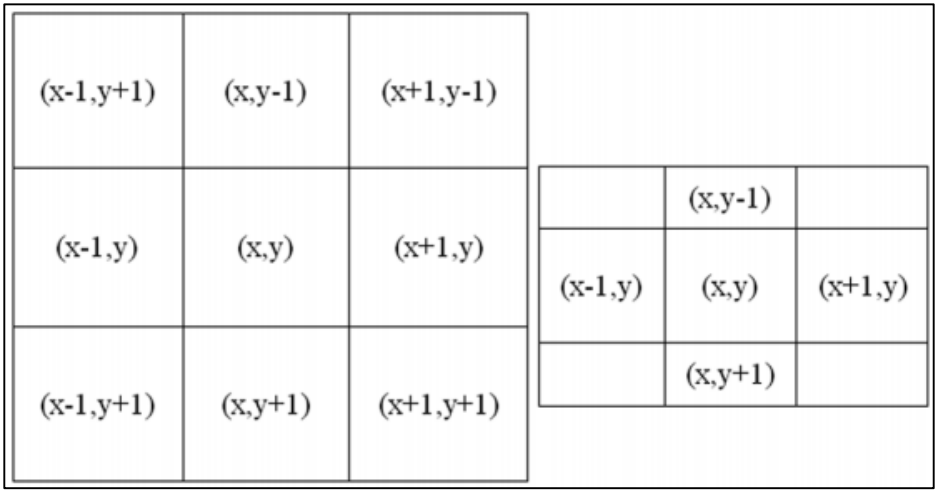

Gambar-41 Sistem koordinat 4 tetangga dan 8 tetangga

Dengan menggunakan metode tersebut, orientasi sudut serat pada benang tekstur dapat diukur secara akurat, kemudian ditentukan nilai index orientasi serat pada benang dengan menggunakan persamaan

$$F = 1(3/2)\, Sin^2 \emptyset \qquad (10)$$

$\emptyset$ merupakan besarnya sudut kemiringan rata-rata dari serat pada benang dan $F$ adalah besarnya index orientasi serat pada benang tekstur.

Hasil pengukuran yang telah dilakukan dengan menggunakan algoritma tersebut dapat dilihat pada Tabel-9. Berdasarkan data yang didapatkan, semakin besar antihan yang diberikan pada benang tekstur, maka akan semakin besar juga derajat $\emptyset$ serat dan orientasi index akan semakin kecil. Berdasarkan analisis menggunakan interval kepercayaan 95%, telah ditemukan bahwa kelima sampel benang memiliki sudut antihan yang berbeda. Tabel-10 menunjukan uji pengukuran sudut dengan menggunakan algoritma yang telah disusun.

Tabel-9 Hasil pengukuran parameter benang tekstur (Ghaderpanah dkk 2014)

| Sample number | Mean of Twist per meter | Density, % | Mean angle of rotation, deg | CV of filament alignment angles | Orientation factor | Crimp concentration, % |
|---|---|---|---|---|---|---|
| 1 | 2165 | 1,55 | 29,70 | 27,34 | 0,55 | 46,8 |
| 2 | 2559 | 1,82 | 31,40 | 26,87 | 0,53 | 51,2 |
| 3 | 2954 | 1,8 | 32,77 | 27,03 | 0,5 | 55,2 |
| 4 | 3189 | 1,87 | 32,97 | 26,94 | 0,5 | 56,4 |
| 5 | 3544 | 1,98 | 32,68 | 26,85 | 0,51 | 56,4 |

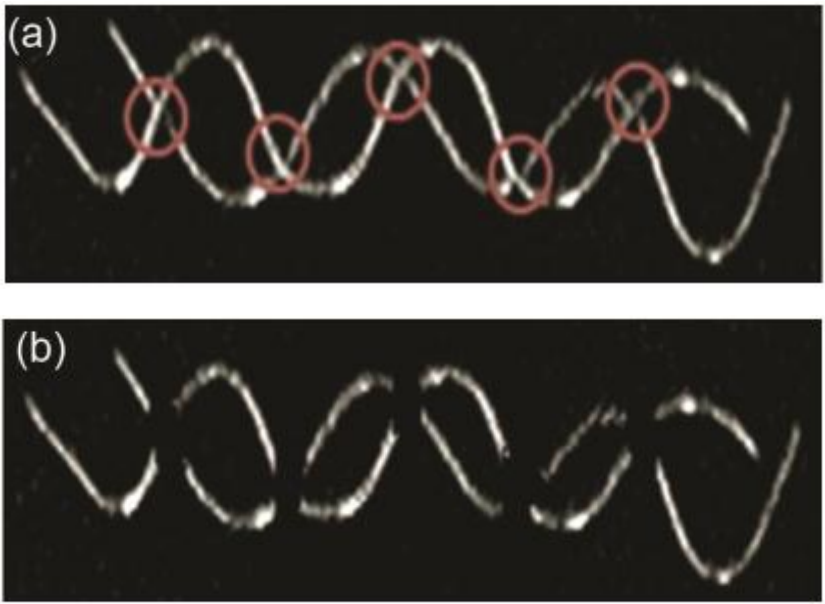

Gambar-42 Gambar simulasi dengan menggunakan *connecting spoints* (a) dan penghilangan titik silang

Tabel-10 Nilai rata-rata sudut berdasarkan hasil pengukuran sampel dan dengan pengukuran *image processing* dari gambar serat

| Gambar hasil *fiber tracing* | Sudut yang diperoleh dari sampel (º) | Sudut yang diperoleh dari algoritma (º) |
|---|---|---|
| 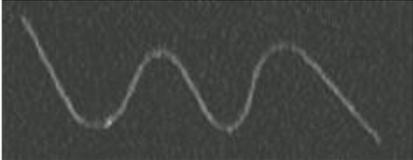 | 55.92 | 56.42 |
| 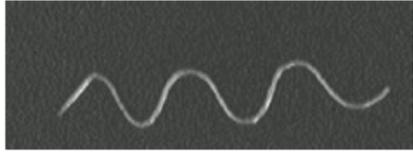 | 44.68 | 45.89 |
| 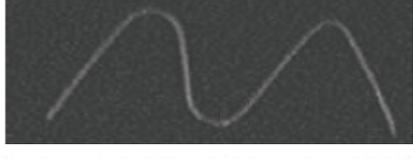 | 61.41 | 61.97 |
| 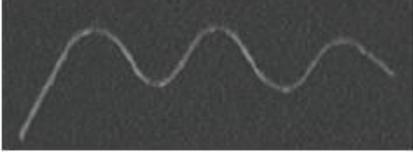 | 49.82 | 50.11 |
| 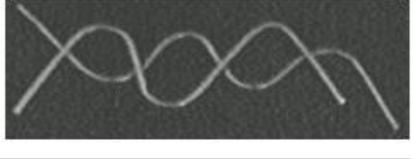 | 63.44 | 63.17 |
| 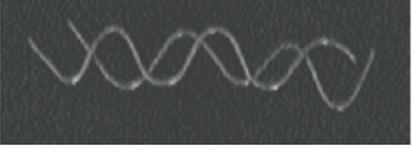 | 44.15 | 45.10 |

Berdasarkan uji pengukuran sudut berbasis algoritma yang dibuat oleh Ghaderpanah, dkk (2014) pada Tabel-10, dapat dilihat bahwa pengukuran sudut secara algoritma tidak berbeda jauh dengan hasil secara manual *selection.*

Pada penelitian Ghaderpanah, dkk (2014), telah ditemukan pula hubungan antara suhu pemanas pada proses *texturizing* dengan orientasi serat pada benang tekstur. Telah ditemukan bahwa semakin tinggi suhu *heater* pada proses *texturizing*, maka serat akan semakin turun derajat orientasinya. Gambar-43 menunjukan grafik pengaruh suhu pemanas terhadap orientasi serat pada proses *texturizing.*

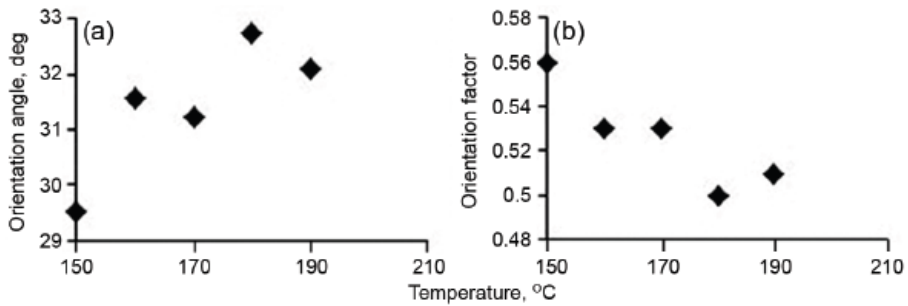

Gambar-43 Pengaruh suhu pemanasan proses *texturizing* terhadap sudut orientasi serat (a) dan terhadap indeks orientasi serat (b)

Tabel-11 Pengaruh *temperature* pada proses *texturizing*

| Sampel | Temperatur pemanas (°C) | Density (%) | Rata-rata sudut orientasi (°C) | CV keteraturan serat | Indeks orientasi | *Crimp concentration (%)* |
|---|---|---|---|---|---|---|
| 1 | 150 | 2.31 | 29.54 | 24.91 | 0.56 | 52.4 |
| 2 | 160 | 2.24 | 31.56 | 25.81 | 0.53 | 55.8 |
| 3 | 170 | 2.01 | 31.23 | 26.77 | 0.53 | 55.9 |
| 4 | 180 | 2.07 | 32.76 | 26.60 | 0.5 | 57.6 |
| 5 | 190 | 2.12 | 32.12 | 26.38 | 0.51 | 56.7 |

Tabel-12 Pengaruh *texturizing speed* pada proses *texturizing*

| Sampel | Kecepatan (m/menit) | Density (%) | Rata-rata sudut orientasi (°C) | CV keteraturan serat | Indeks orientasi | *Crimp concentration (%)* |
|---|---|---|---|---|---|---|
| 1 | 40 | 2.02 | 31.20 | 26.04 | 0.53 | 58.3 |
| 2 | 60 | 1.86 | 30.78 | 26.73 | 0.54 | 57.9 |
| 3 | 80 | 1.98 | 29.93 | 26.03 | 0.55 | 56 |
| 4 | 100 | 1.95 | 29.21 | 26.05 | 0.57 | 53.2 |
| 5 | 120 | 2.15 | 29.04 | 25.82 | 0.57 | 53.2 |

Tabel-13 Uji kesamaan rata-rata sudut orientasi serat pada berbagai nilai *texturizing speed* yang berbeda

| Kecepatan (m/menit) | N | Pengelompokan nilai, dengan α = 0,05 | | |
|---|---|---|---|---|
| | | 1 | 2 | 3 |
| 120 | 150 | 29.0351 | -- | -- |
| 100 | 150 | 29.2099 | -- | -- |
| 80 | 150 | -- | 29.9270 | -- |
| 60 | 150 | -- | -- | 30.7812 |
| 40 | 150 | -- | -- | 31.2031 |
| Sig. | -- | 0.617 | 1.000 | 0.228 |

Berdasarkan Tabel-12 dan Tabel-13, dapat dilihat bahwa kecepatan *texturizing* pada produksi benang tekstur dapat mempengaruhi sudut orientasi serat pada benang. Hal tersebut diperkuat dengan hasil analisis uji kesamaan pada Tabel-13, dengan nilai kepercayaan 95%, telah didapatkan bahwa kecepatan berpengaruh terhadap nilai orientasi serat.

# 9. PENILAIAN KENAMPAKAN BENANG (GRADE BENANG) BERBASIS IMAGE PROCESSING

Menurut Li dkk (2012), Penilaian *grade* benang sering disebut juga sebagai penilaian mutu kenampakan permukaan benang, adalah salah satu prosedur pengujian penting di industri tekstil. Pada umumnya, proses inspeksi mutu kenampakan benang dilakukan dengan melakukan pengamatan secara langsung yang dilakukan oleh seorang yang ahli dalam membandingkan kenampakan benang dengan foto standar mutu benang.

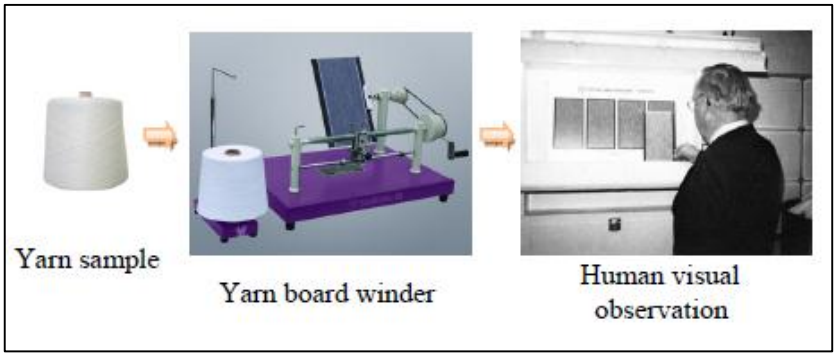

Gambar-44 Proses penilaian kenampakan benang dengan metoda konvensional

Pada metoda yang telah ditemukan oleh Li dkk (2012), penentuan *grade* benang dilakukan dengan menggunakan prinsip pengolahan citra. Gambar-45 menunjukan proses penentuan *grade* benang secara *image processing.*

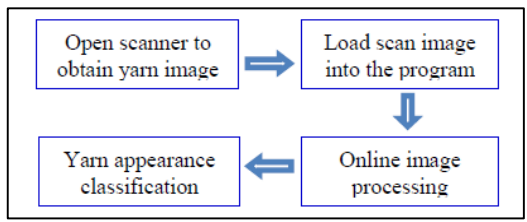

Gambar-45 Proses penentuan grade benang dengan metode pengolahan citra

Li dkk (2012) telah berhasil mengembangkan suatu perangkat lunak yang dapat menilai mutu kenampakan benang berbasis *image processing*. Gambar-39 menunjukan tampilan perangkat lunak penilai *grade* benang yang telah dirancang.

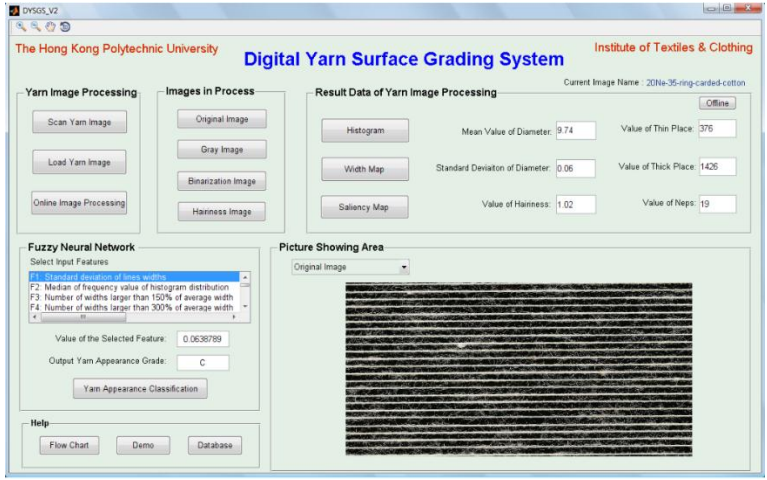

Gambar-46 Perangkat lunak penilai *grade* benang berbasis *image processing*

Pada bagian antar muka perangkat lunak, tombol "Scan Yarn" berfungsi untuk membuka pemindai benang. Langkah pengujian *grade* kenampakan benang berbasis *image processing* Li dkk (2012) dapat dilhat pada Gambar-47. Hasil pemindaian yang lebih detail akan memberikan hasil pengujian yang lebih akurat.

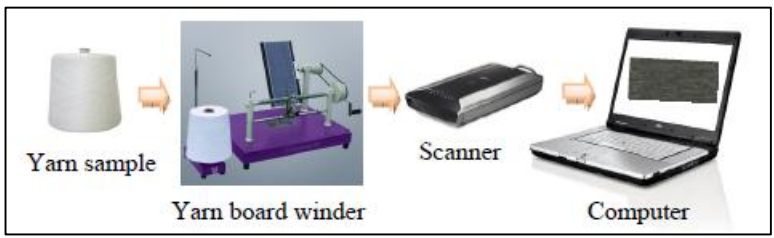

Gambar-47 Tata cara pengujian *grade* kenampakan benang secara pengolah citra

Gambar-48 menunjukkan dua gambar pindaian dari dua benang yang berasal dari kelas mutu yang berbeda. Li dkk (2012) telah menggunakan kedua sampel ini untuk menunjukkan kinerja metodologi dalam sistem saat menganalisis benang dengan kelas yang berbeda.

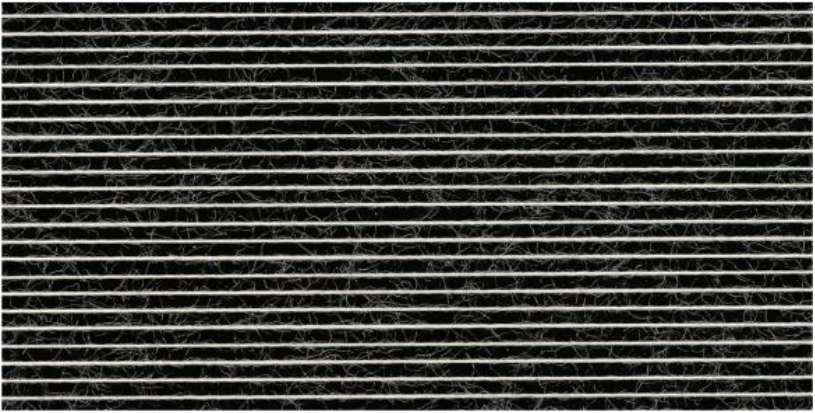

(a) Benang *grade* A

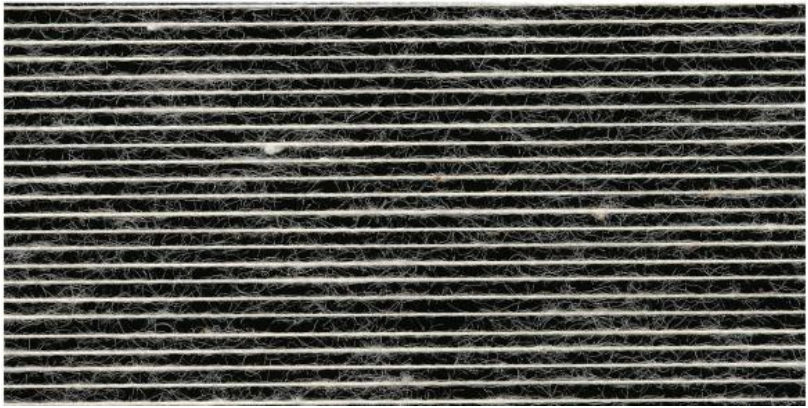

(b) Benang *grade* B

Gambar-48 Gambar pindaian dari benang *grade* A (a) dan pindaian dari benang *grade* B (b)

Secara garis besar, proses pengolahan citra terbagi atas dua buah proses, yaitu proses *preprocessing* dan proses *data extraction*. Proses preprocessing meliputi proses transformasi *wavelet* dan mengubah citra dalam bentuk citra biner. Transformasi wavelet merupakan analisis multi-resolusi yang digunakan untuk memisahkan bagian benang dengan bagian bulu atau serat pada permukaan benang. Gambar-49 menunjukan hasil analisis wavelet citra digital benang pada Gambar-48.

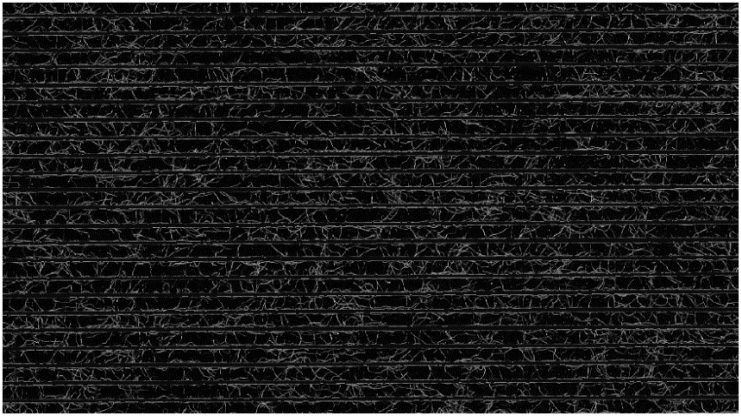

(a) Benang *grade* A

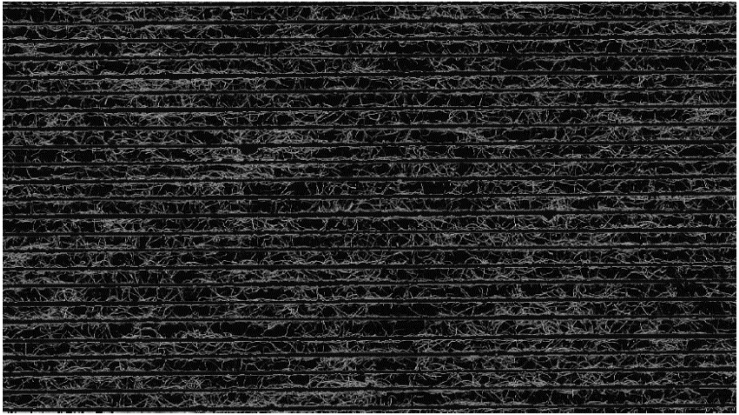

(b) Benang *grade* B

Gambar-49 Hasil analisis wavelet

Setelah proses transformasi wavelet, citra digital kemudian diubah dalam bentuk citra biner. Hasil pengubahan citra hasil analisis wavelet menjadi citra biner dapat dilihat pada Gambar-50. Hal ini dapat mempermudah analisis *fast Fourier Transform* (FFT) proses *tresholding* dengan metode OTSU.

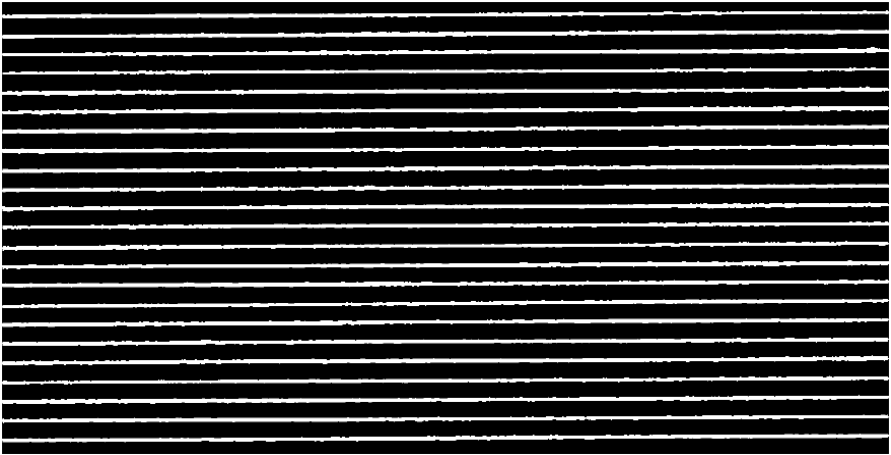

(a) Benang *grade* A

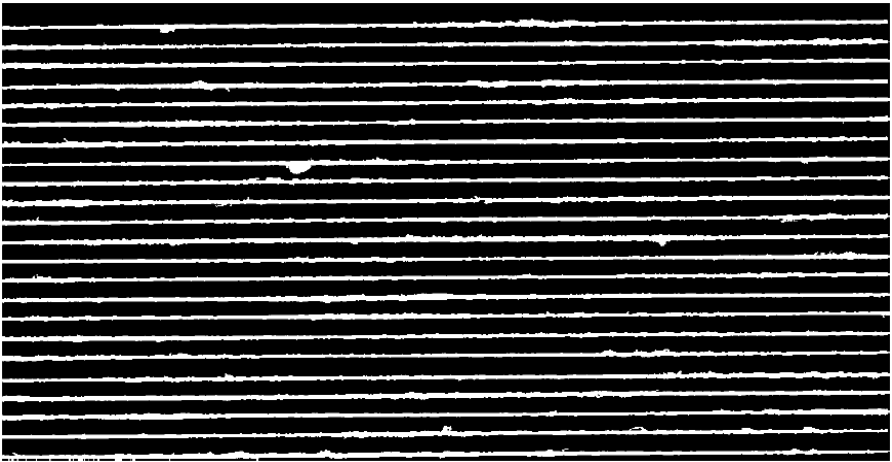

(b) Benang *grade* B
Gambar-50 Citra biner hasil analisis wavelet

Berdasarkan citra biner pada Gambar-50, Li dkk (2012) melakukan analisis pengukuran diameter benang dengan menggunakan metode histogram. Gambar-51 dan Gambar-52 menunjukan histogram dari nilai *pixel-pixel* yang ada pada citra Gambar-50 untuk mengidentifikasi diameter benang.

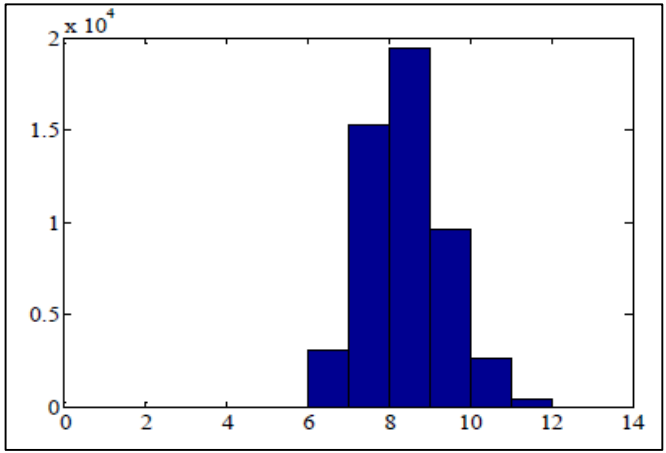

(a) Benang *grade* A

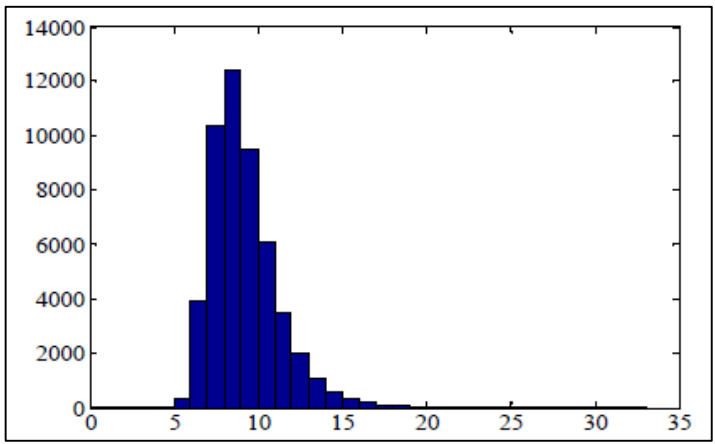

(b) Benang *grade* B

Gambar-51 Grafik histogram diameter benang

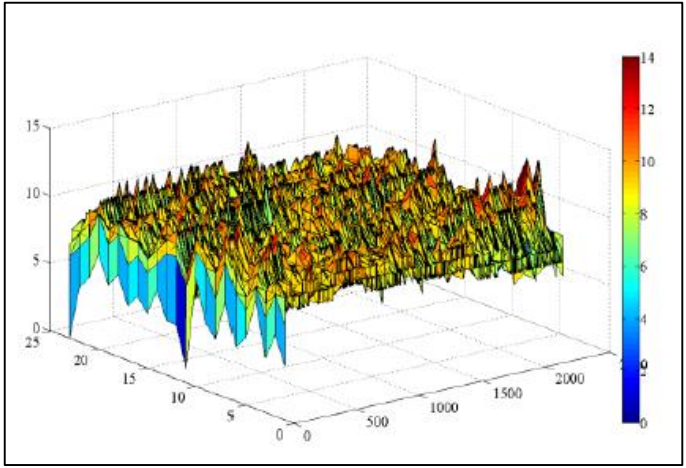

(a) Benang *grade* A

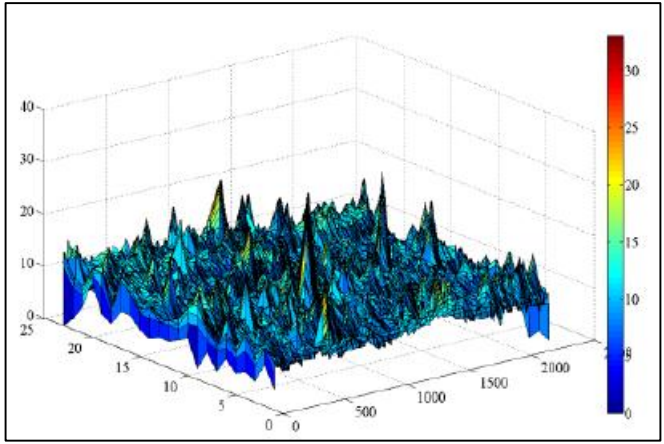

(b) Benang *grade* B
Gambar-52 Grafik *width map* diameter benang

Prinsip pengambilan nilai pada histogram adalah dengan melihat nilai *color value* pada *pixel-pixel*. Grafik pada Gambar-51 akan menghitung jumlah setiap *pixel* yang memiliki nilai yang sama, kemudian dibentuk ke dalam distribusi kumulatif. Puncak pada grafik akan merepresentasikan benang pada kain, hal tersebut disebabkan karena benang digambarkan sebagai daerah berwarna putih pada gambar, sehingga nilai *color value*-nya akan lebih besar.

*Artificial Neural Network* telah digunakan untuk mendeteksi cacat yang ada pada struktur benang. Berdasarkan pengujian kenampakan benang yang telah dilakukan, pada berbagai jenis benang dengan nomor benang yang bervariasi ($Ne_1$ 20 – $Ne_1$ 80), hasilnya menunjukan bahwa sekitar 87% dari 170 sampel dapat secara benar ditentukan *grade*-nya oleh *artificial neural network*. Metode tersebut bekerja atas dasar grafik *saliency map* yang telah dihasilkan berdasarkan struktur benang, *saliency map* dari citra Gambar-48 dapat dilihat pada Gambar-53.

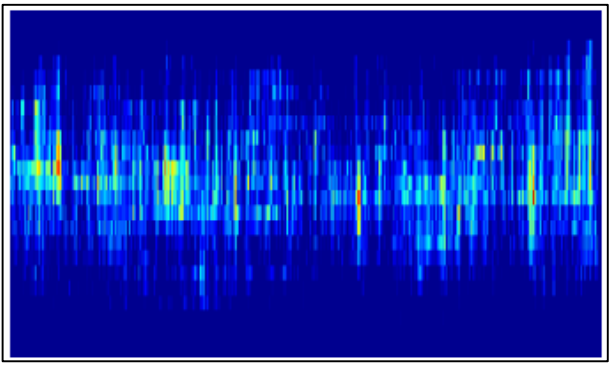

(a) Benang *grade* A

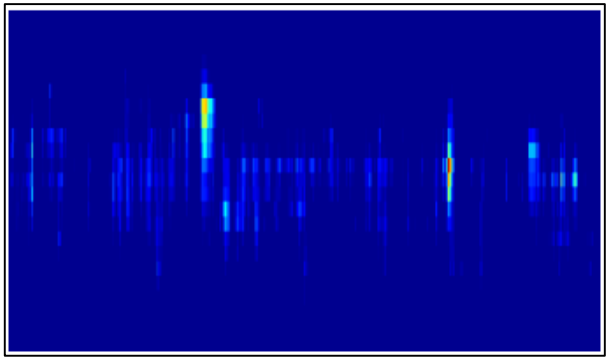

(a) Benang *grade* B
Gambar-52 Grafik *width map* diameter benang

Sehingga secara garis besar, Li dkk (2012) mengatakan bahwa sistem yang telah dikembangkan sudah dapat memberikan hasil pengukuran *grade* kenampakan benang secara akurat.

## 10. DAFTAR PUSTAKA